\documentclass{aa}
\usepackage{graphicx}
\usepackage{longtable}
\usepackage{natbib} 
\bibpunct{(}{)}{;}{a}{}{,}
\usepackage{txfonts}
\usepackage{siunitx}
\usepackage{enumitem}
\usepackage{float}
\usepackage{placeins}
\usepackage{float}
\usepackage{mwe,tikz}
\usepackage[colorlinks=true,linkcolor=blue,citecolor=blue,urlcolor=blue]{hyperref}
 
\usepackage[x11names]{xcolor}

\newcommand{\msunpyr}{\ifmmode{\,M_{\odot}\,\mbox{yr}^{-1}} \else{ M$_{\odot}$/yr}\fi}

\newcommand{\kms}{\ifmmode{\,\mbox{km}\,\mbox{s}^{-1}}\else{km\,s^{-1}}\fi}
\newcommand{\kpc}{\ifmmode {\,\mbox{kpc}} \else{kpc}\fi}
\newcommand{\msun}{\ifmmode M_{\odot} \else M$_{\odot}$\fi}
\newcommand{\rsun}{\ifmmode R_{\odot} \else R$_{\odot}$\fi}
\newcommand{\lsun}{\ifmmode L_{\odot} \else L$_{\odot}$\fi}
\newcommand{\zsun}{\ifmmode Z_{\odot} \else $Z_{\odot}$\fi}
\newcommand{\xsun}{\ifmmode X_{\odot} \else $X_{\odot}$\fi}
\newcommand{\velo}{\ifmmode\varv\else$\varv$\fi}
\newcommand{\vinf}{\ifmmode\velo_\infty\else$\velo_\infty$\fi}
\newcommand{\rgal}{\ifmmode \,R_{\mathrm{gal}} \else R$_{\mathrm{gal}}$\fi}

\begin{document}

 	\title{The drastic impact of Eddington-limit induced mass ejections on massive star populations}
 	
 	\subtitle{}
 	\titlerunning{The drastic impact of Eddington-limit induced mass ejections on massive star populations}
 	
 	\author{D.~Pauli\inst{\ref{IvS}} \and N.~Langer\inst{\ref{Aifa},\ref{MPIFR}} \and A. Schootemeijer\inst{\ref{Aifa}} \and P. Marchant\inst{\ref{Gent}} \and H. Jin\inst{\ref{Aifa}} \and A. Ercolino\inst{\ref{Aifa}} \and A.~Picco\inst{\ref{IvS}} \and R.~Willcox\inst{\ref{IvS}} \and H. Sana\inst{\ref{IvS},\ref{LGI}}
        }
 	
 	\authorrunning{D. Pauli et al.}
 	
 	\institute{Institute of Astronomy, KU Leuven, Celestijnenlaan 200D, 3001 Leuven, Belgium\label{IvS} \and 
                Argelander Institut für Astronomie, Auf dem Hügel 71, DE-53121 Bonn, Germany\label{Aifa} \and
                Max-Planck-Institut für Radioastronomie, Auf dem Hügel 69, DE-53121 Bonn, Germany\label{MPIFR}\and
                Sterrenkundig Observatorium, Universiteit Gent, Gent, Belgium\label{Gent} \and
                Leuven Gravity Institute, KU Leuven, Celestijnenlaan 200D, box 2415, 3001 Leuven, Belgium\label{LGI}
 	}
 	
 	\date{Received ; Accepted}
 	
 	\abstract
        {
        Massive stars emit copious amounts of radiation, profoundly affecting their environment in galaxies and contributing to the reionization of the Universe. However, their evolution and thus their ionizing feedback are still not fully understood. One of the largest gaps in current stellar evolution calculations is the lack of a model for the mass ejections that occur when the stars reach the Eddington limit, such as during a Luminous Blue Variable (LBV) phase.
        }
        {
        Here, we aim to remedy this situation by providing a physically motivated and empirically calibrated method applicable in any 1D stellar evolution code to approximate the effect of such mass loss on stellar evolution.
        }
        {
        We employed the 1D stellar evolution code MESA, in which we implement a new mass-loss prescription that becomes active when stellar models inflate too much when reaching the Eddington limit. We used lines of constant inflation factors in the Hertzsprung-Russell diagram (HRD) for a simple empirical calibration of the threshold value. We calculated synthetic massive-star stellar populations using grids of single-star models with this mass loss prescription compared them with the observed populations in the Large and Small Magellanic Clouds. Further, with already computed grids of binary evolution models, we investigated the impact of binarity on our predictions.
        }
        {
        Our single-star models reproduce key features of the observed stellar populations, namely, (i) the absence of stars located beyond the Humphreys-Davidson limit; (ii) an upper limit of red supergiant (RSG) luminosities; (iii) the faintest observed single Wolf-Rayet (WR) stars; (iv) the absolute number of O-stars, WRs, and RSGs; (v) WO stars in low metallicity environments; and (vi) the positions of LBV stars in the HRD. We show that binarity still plays an important role in explaining the observed WR stars. However, a large fraction of the binary population can also be explained via self-stripping. At the same time, our binary population explains the 70\% binary fraction of O-stars and the 40\% binary fraction of WR stars. However, our synthetic population also has caveats, such as an overproduction of bright H-free WN stars.
        }
        {
        Our results show that the effect of the Eddington-limit induced mass ejections on the structure and evolution of massive stars can remove the tension between predicted and observed massive star populations. A more fundamental treatment of these effects, particularly for hydrogen-poor stars, is needed to fully comprehend massive star evolution.
        }
 	
 	\maketitle

 	\section{Introduction} 
 	\label{sec:intro} 
        Massive stars ($M_\mathrm{ini}\gtrsim8\,\msun$), although relatively rare, are the powerful engines of the Universe. Through their intense ionizing radiation, strong stellar winds, Luminous Blue Variable (LBV) eruptions, and supernova explosions, they shape their surroundings and are the regulators of star formation and sources of chemical enrichment within their host galaxies \citep[e.g.,][]{mae1:83,dra1:03,hop1:14,ram1:18,cro1:19}. Despite their significance, the evolution of massive stars, particularly at the high-mass end, where their impact is the strongest, remains poorly understood. The predicted evolution of these stars is highly dependent on the assumptions about internal mixing processes and mass-loss rates applied during different evolutionary stages \citep[e.g.,][]{she1:20,gil1:21,sab2:22,zap1:24}.

        One of the major challenges facing current stellar evolution models is their tendency to overproduce extremely luminous red supergiant (RSG) stars \citep[e.g.,][]{bro1:11,geo1:13,cho1:16,lim1:18,cos1:25} beyond the empirical Humphreys-Davidson (HD) limit ($\log L/\lsun \gtrsim 5.5$), where no stars are observed \citep{hum1:79,dav1:18}. This issue has been the subject of numerous studies exploring the effects of mixing processes and different RSG mass-loss rates on the synthetic RSG populations across various metallicities \citep{sch1:18, kle1:20, hig1:20, zap1:24,dec1:24}. However, a solution to this problem remains elusive.

        Luminous Blue Variables are massive stars that have evolved into a region in the Hertzsprung-Russell diagram (HRD) where their outer atmosphere is barely gravitationally bound, leading to enhanced mass-loss rates with possibly episodic outbursts. During the LBV phase, a star can lose several solar masses, making it a critical component of massive stellar evolution that could potentially alleviate some of the tension of the overluminous RSG problem. However, our understanding of the LBV phase is limited. Its impact on the evolution of massive stars is unclear \citep{smi1:26}, and due to the absence of a consistent framework for modeling LBV eruptions, it is often ignored in detailed stellar evolution calculations. Early pioneering modeling attempts tried to incorporate the effect of LBV eruptions in stellar evolution calculations, by increasing the mass-loss rates of stars massive enough to undergo an LBV phase as soon as they enter a specific evolutionary stage (such as the post-main sequence, \citealt{van1:91}) or to enhance the mass-loss in regions unstable against strange-mode pulsations \citep{lan1:94}. In recent years, progress has been made in understanding the physics, impact, and modeling of LBV mass ejections \citep{gra1:21,muk1:24,che1:24}. Nevertheless, a physical model that prevents stellar evolution simulations from surpassing the HD limit for a significant duration, especially at low metallicity ($Z$), is still lacking.

        A third major issue is the difficulty to explain the existence of the least luminous, apparently single WR stars, particularly in metal-poor galaxies \citep{she1:20}. The strength of UV radiation-driven stellar winds weakens \citep[e.g.,][]{vin1:01,mok2:07,hai1:17,pau1:25}, which leads to the expectation that the minimum luminosity of WR stars formed by single stars should increase with decreasing $Z$. Consequently, stellar evolution models predict that the faintest WR stars in low-metallicity environments are all post-interaction binary products \citep[e.g.,][]{smi1:14, mas1:21, pau1:22}. Contrarily, dedicated multi-epoch surveys of the least-luminous WR stars in low-$Z$ environments strongly suggest that these stars are currently single \citep{foe1:03, sch1:24}. To explain the presence of these faint single WR stars, single-star stellar evolution calculations require enhanced mass-loss rates \citep[e.g.,][]{yoo1:17, sab1:22}. Yet this approach is in contrast to empirical mass-loss measurements, which over the past few decades have exposed lower mass-loss rates for hot stars than those commonly assumed in standard mass-loss prescriptions \citep{hai1:17, she1:19, pau1:25}. 

        The Large (LMC) and Small Magellanic Clouds (SMC) are two nearby galaxies with low metallicities (i.e., Fe-group elements) of $Z_\mathrm{LMC}=0.5\,\zsun$ and $Z_\mathrm{SMC}=0.14\,\zsun$, respectively \citep{vin2:23}. These galaxies have been subject to numerous observing campaigns, and their massive star content is mostly known. In particular, their RSG populations \citep{dav1:18,yan1:23} as well as the WR population, including their binary properties \citep{neu1:18,bar1:01,bar2:01,foe1:03,hai1:14,hai1:15,she1:19}, are mostly known. This makes these galaxies ideal test beds to benchmark stellar evolution models.

        In this work, we present a new physically motivated and empirically calibrated method for modeling Eddington-limit induced mass ejections in 1D stellar evolution codes. We investigate how these mass ejections influence massive star populations. We compare the resulting synthetic populations with the observed massive star populations of the LMC and SMC.

        Section~\ref{sec:Methods} outlines the physical assumptions incorporated into our stellar evolution calculations, the new model for the Eddington-limit induced mass ejections, and the assumptions used to generate the synthetic stellar populations. The resulting synthetic populations are presented in Sect.~\ref{sec:result}. Section~\ref{sec:discuss} discusses their sensitivity to the assumptions made during the modeling process as well as their implications for understanding stellar populations. Our results are summarized in Sect.~\ref{sec:conclusions}.
        
 	\section{Methods}
    \label{sec:Methods}

    \subsection{Stellar evolution modeling}
    \label{sec:setup}

    We employed the Modules for Experiments in Stellar Astrophysics (MESA) code, version 24.08.1 \citep{pax1:11,pax1:13,pax1:15,pax1:18,pax1:19,jer1:23}, to calculate stellar evolution models at LMC ($Z_\mathrm{LMC}=\num{6.17e-3}$) and SMC ($Z_\mathrm{SMC}=\num{2.44e-3}$) metallicity. Our models cover initial masses spanning from $M_\mathrm{ini}\approx10\,\msun\,\text{--}\,400\,\msun$ (${\log(M_\mathrm{ini}/\msun)=\SIrange{1}{2.65}{}}$) in steps of $\Delta\log(M_\mathrm{ini}/\msun)=0.05$. For simplicity, we assume that all stars are born with a moderate rotation velocity of $\varv_\mathrm{rot,\,ini}=\SI{100}{km\,s^{-1}}$ (i.e., where the observed $\varv_\mathrm{rot}$-distribution peaks, e.g., see \citealt{ram1:15,ram1:19}). In total, 33 single-star models per initial metallicity and per input physics variation (see Sect.~\ref{sec:models}) are calculated.

    A detailed list of the input physics is provided in Appendix~\ref{app:MESA}, and our source files can be downloaded from Zenodo.\footnote{\url{https://doi.org/10.5281/zenodo.17397155}} Here, only the most relevant physics are mentioned. Our models incorporate semiconvective mixing with $\alpha_\mathrm{sc}=1$ \citep{lan1:83, sch1:19}, and step core overshooting with $f_\mathrm{ov}=0.345$ and $f_{\mathrm{ov},\,0}=0.01$ \citep{bro1:11} during core hydrogen and helium burning. To account for stabilizing chemical gradients that may limit the growth of the overshooting regions, we include the Brunt-Vaisala frequency and set the stabilizing Brunt composition gradient to $B=0.1$. As a result, in most of our calculations, overshooting during core helium burning occurs primarily in the very early stages, before a steep chemical gradient between the convective core and the envelope forms. 
    
    During the late evolutionary stages of the most massive stars ($M_\mathrm{ini}\gtrsim40\,\msun$), with helium core masses exceeding $13\,\msun$, we encountered convergence issues. To address this, we employed for these models after core hydrogen depletion the superadiabatic reduction method of MESA \citep[][Section 7.2]{jer1:23}. This method enhances energy transport locally in layers close to the Eddington limit, resulting in less impact on the final stellar structure compared to MLT++. For the superadiabatic reduction, we set the energy transport enhancement factors to ${\alpha_1=\alpha_2=3.5}$, which allows our models to converge within a reasonable time frame while minimizing deviations from standard mixing length theory. Since this option is only turned on after core hydrogen depletion, all of our core hydrogen burning models approaching the Eddington limit locally are affected by envelope inflation \citep{san1:15}.

    In our standard setup, we account for stellar winds as follows. For hot stars with radiative envelopes, we apply the recent empirical mass-loss recipe from \citet{pau1:25}, which depends on the classical Eddington factor $\Gamma_{\!\text{e}}$ and the metallicity $Z$. However, this approximation is only valid for optically thin winds, and it breaks down for stars approaching the Eddington limit. Therefore, when $\log\Gamma_{\!\text{e}} > -0.4$, we switch to the empirical Wolf-Rayet (WR) mass-loss recipe from \citet{she2:19} if the mass-loss rate is higher than the one from \citet{pau1:25}. To avoid abrupt changes in the mass-loss rates that could lead to numerical issues, we linearly interpolate between the hot star and WR mass-loss recipes in the range from $\log\Gamma_{\!\text{e}}=-0.5\,\,\text{to}\,\,-0.4$. For RSGs, we adopt the empirical mass-loss recipe from \citet{yan1:23}, which has been shown to best reproduce the RSG population in the SMC \citep{zap1:24}. While the launching mechanism of RSG winds remains a topic of debate, recent theoretical models suggest that these winds are powered by turbulence near the stellar surface and are therefore $Z$-independent \citep{kee1:21}. This is further supported by observational evidence, which shows that even at extremely low $Z$, no RSG with $\log(L/\lsun)\gtrsim5.5$ is observed \citep{dav1:18,mcd1:22,sch2:25}.
    To switch to an RSG wind in our models, we require that the layers where hydrogen can recombine ($T < \SI{10000}{K}$) are convective and account for more than $0.1\%$ of the total hydrogen envelope mass. Further, we account for Eddington-limit induced mass ejections as associated with LBVs or the radial pulsations of hydrodynamically unstable RSG envelopes when approaching the Eddington limit \citep[e.g.,][]{heg1:97,yoo1:10,bro1:25,lap1:25}. 
    
    \subsection{Eddington-limit induced mass ejections}
    
    \begin{figure*}[tbhp]
        \centering
        \includegraphics[width=\textwidth]{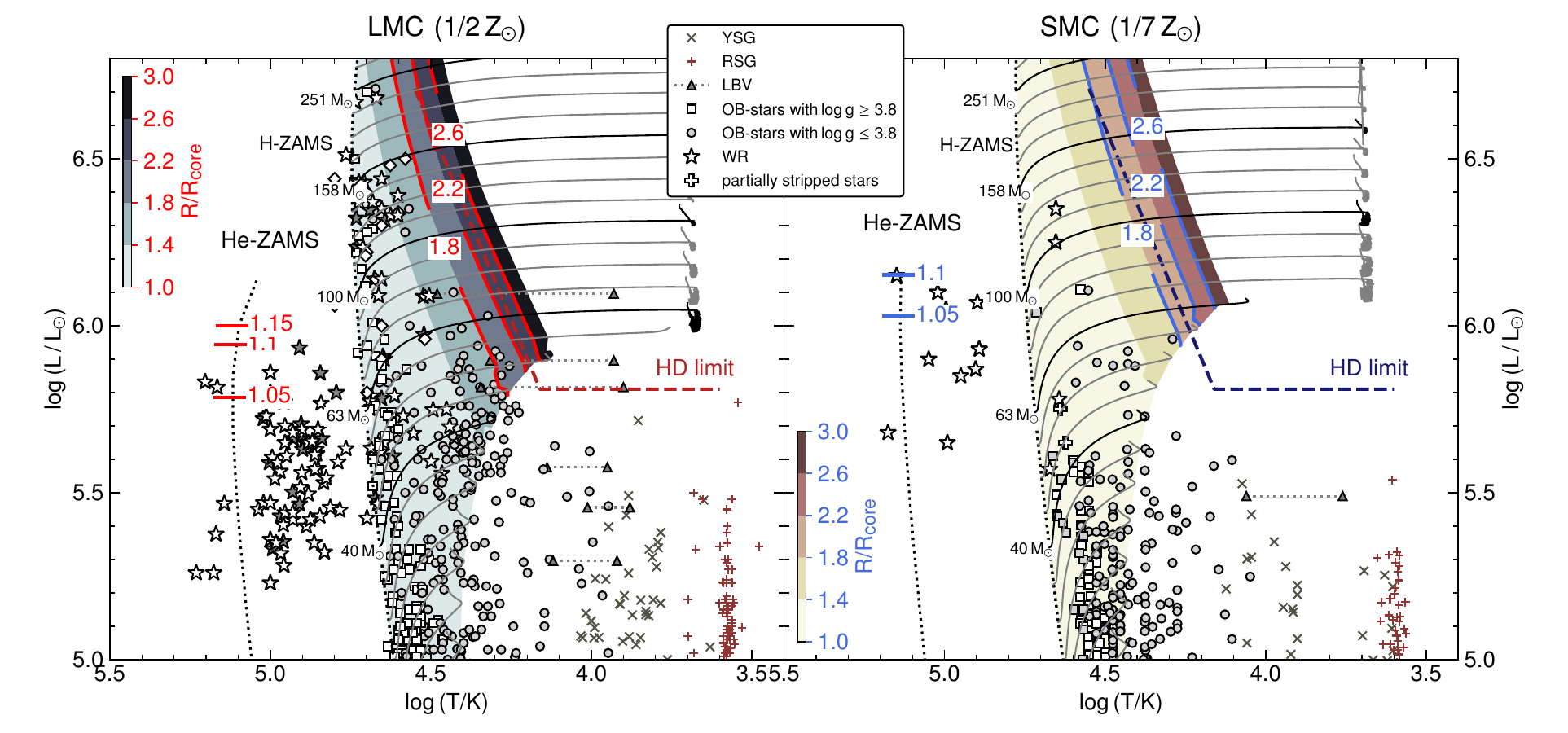}
        \caption{Hertzsprung-Russell diagrams of the massive star population (symbols) in the LMC (left) and SMC (right). The different symbols mark the positions of observed RSG \citep[small red plusses,][]{dav1:18,yan1:23}, YSG \citep[crosses,][]{neu1:10,neu1:12}, LBVs \citep[filled triangles with quiescent and outburst positions,][]{hum1:16,kal1:18}, OB stars (open circles and squares, \citealt{eva1:04,tru1:04,tru1:05,hun1:08,bes1:11,bou1:13,urb1:17,cas1:18,sch2:18,ram1:18,ram1:19,duf1:19,bes1:20,bou1:21,ric1:22,pau1:23,ber1:24,gom1:25,alk1:25}), partially stripped stars \citep[white large plusses][]{pau1:22,ric1:23,ram1:23,ram1:24}, and WR stars \citep[white stars for apparent single, gray stars for binaries][]{cro1:02,hai1:14,hai1:15,tra1:15,she1:16,she1:19}. The empirical HD limit \citep{hum1:79} is indicated by the hatched regions. Solid black and gray lines show stellar evolution tracks of the Model no-Eject until core hydrogen depletion. For clarity only the tracks with initial masses $M_\mathrm{ini}=40\,\msun$, $63\,\msun$, $100\,\msun$, $158\,\msun$, and $251\,\msun$ are labeled and colored in black. Dotted black lines mark the H-ZAMS and He-ZAMS. Contours in the background highlight the degree of inflation covering $R/R_\mathrm{core}=1\,\text{--}\,3$. Inflations of $R/R_\mathrm{core}=1.6$, $2.2$, and $2.8$ for stars during hydrogen burning, and $R/R_\mathrm{core}=1.05$, $1.1$, and $1.15$ for the He-ZAMS models are marked by colored lines and numbers.}
        \label{fig:hrd_inflation}
    \end{figure*}

    A connection of the LBV phenomenon with the Eddington limit is evident since the pioneering work of \citet{lam1:88} and \citet{ulm1:88}, who showed that the empirical HD-limit is essentially recovered by the Eddington limit in stellar model atmospheres. This connection is strengthened by 1D stellar structure and evolution models \citep{ish1:99,gra1:12,koe1:15,san1:15,gra1:21}, which found that stars approaching their Eddington limit locally in their envelopes inflate their radius to allow for a sufficient radiative energy transport, a phenomenon typical for hydrostatic codes that cannot be recovered by 3D radiation-hydrodynamic calculations \citep{jia1:15,jia1:18,pon1:21}.
 
    In contrast to \citet{che1:24}, who model eruptive mass loss in 1D massive star models based on the local super-Eddington luminosity, we choose an empirically calibrated estimate of the global instability of the inflated stellar envelopes. In the absence of systematic stability analyses of such envelopes, we assume here that a stronger inflation leads to a higher chance for instability. A possible instability mechanism is provided by strange mode pulsations \citep{gau1:90} which have been shown to occur in supergiant stellar models near the Eddington limit \citep{kir1:93,san2:13,lov1:14}, and which have growth time scales that are close to the hydrodynamic time scale of the envelope. To approximate Eddington-limit induced mass ejections, we assume that a star can inflate to a certain degree without becoming unstable, but that there must be a threshold value after which a star will expel its envelope until it returns to a ``stable'' inflation (i.e., when inflation is again below the given threshold). In our models, we fully account for the effect of envelope inflation during core hydrogen burning. During core helium burning, we apply the local superadiabatic reduction method from MESA for numerical stability, which still allows for significant inflation. 

    To determine the extent of inflation $R/R_\mathrm{core}$ in a stellar evolution model with radius $R$, we need to compute the radius that the star would have if inflation did not occur, which we designate $R_\mathrm{core}$. This radius can technically be recovered by artificially increasing the efficiency of convective energy transport, for instance, by increasing the mixing-length parameter \citep{san1:15,nat1:25}. However, this is computationally expensive. Instead, we follow \citet{san1:15}, who showed that $R_\mathrm{core}$ can be well approximated by the location in the stellar envelope at which the gas pressure ($P_\mathrm{gas}$) contribution to the total pressure ($P$) drops below $P_\mathrm{gas}/P \simeq 0.15$. As shown by \citet{san1:15}, the result is not sensitive on the threshold value. Increasing the threshold to $0.30$ impacts $R_\mathrm{core}$ by less than $5\%$. We confirm that this also holds for our model calculations.   

    To calibrate the extent of inflation beyond which Eddington-limit induced mass ejections should be triggered, we calculated stellar evolution models up to core hydrogen depletion using the physical setup described in Sect.~\ref{sec:setup}. In Fig.~\ref{fig:hrd_inflation}, we compare the model tracks to the observed stellar populations in the HRD, for the LMC and the SMC. Contours indicate the amount of inflation in our models across the HRD. Interestingly, the slope of lines with fixed inflation values is parallel to the HD limit, supporting our hypothesis that a certain amount of inflation in the stellar envelope might be stable. One can see that core hydrogen-burning stars exceed the HD limit when their inflation reaches $R/R_\mathrm{core} \approx 2$ \citep[see also][]{san1:17}, which we adopt as the triggering criterion for Eddington-limit induced mass ejections for such stars.

    Since it is uncertain whether this criterion would also hold for helium stars, such as the classical WR stars, and because for those, we need to employ the superadiabatic reduction method from MESA, we also calculated the extent of inflation for stars on the He-ZAMS, which we indicate in Fig.~\ref{fig:hrd_inflation}.  From inspection, it is clear that, regardless of metallicity, the maximum inflation for helium stars appears to be $R/R_\mathrm{core} = 1.1$, as no classical WR stars are observed above this threshold (see also figure 18 of \citet{koe1:15}). 
    
    Based on these empirical limits, we defined the maximum allowed inflation in our models as 
    \begin{equation}
        R_\mathrm{max}/R_\mathrm{core} = 2\,\alpha(X_\mathrm{H}),
    \end{equation}
    with a scaling factor, $\alpha$, depending on the surface hydrogen abundance, $X_\mathrm{H}$, as
    \begin{equation}
        \alpha(X_\mathrm{H}) = \begin{cases}
            1 & X_\mathrm{H}>=0.2\\
            2.125\,X_\mathrm{H}+0.575 & X_\mathrm{H}<0.2   .
        \end{cases}
    \end{equation}
    As shown by \citet{pet1:06}, mass loss with rates above a critical value can reduce the size of the inflated envelopes. Therefore, to approximate the mass lost during an ejection episode, we adopted the critical mass-loss rate ($\dot{M}_\mathrm{eject}$) of \citet{pet1:06} needed to prevent a star from further inflating as 
    \begin{equation}
        \dot{M}_\mathrm{eject} = 4\pi\,R_\mathrm{core}^2\,\rho_\mathrm{min}\,\sqrt{\dfrac{\mathrm{G}M}{R_\mathrm{core}}},
    \end{equation}
    where $M$ is the total mass of the star and $\rho_\mathrm{min}$ is the minimum density in the inflated layers. To avoid extreme jumps in the mass-loss rates, we increased the mass-loss rate linearly between ${R/R_\mathrm{core}=\alpha(X_\mathrm{H})\cdot1.9}$ and $R_\mathrm{max}/R_\mathrm{core}$. For illustrative purposes, we display in Fig.~\ref{fig:hrd_phases} the stellar evolution tracks calculated with the Eddington-limit induced mass ejections and indicate their ejection episodes. One can see that the starting points of the ejection episode in the models overlap with the observed LBVs in the LMC and SMC.

    We performed test calculations where we varied the inflation threshold for the maximum inflation by $50\%$. While this slightly alters the number of models that exceed the HD limit and the maximum WR luminosity, it does not significantly affect the total time spent during the OB or WR phase of a specific model. This gave us confidence that our chosen approach allows for the impact of mass ejections, such as during an LBV phase, on stellar populations to be studied without being overly sensitive to the choice of the threshold parameters.

    We also note that the critical mass-loss rate, and thus a star's subsequent evolution, is only weakly sensitive to the choice of stellar-wind prescription during hot-star evolutionary phases, since typical variations among commonly used OB- or WR-star mass-loss recipes primarily affect the integrated mass loss but are not high enough to significantly alter the local density structure and thus prevent the effect of envelope inflation, especially at low metallicity. However, when adopting different mass-loss prescriptions, the empirical limits used to define the maximum allowed degree of inflation may need to be recalibrated.

    \subsection{Model variations}
    \label{sec:models}
    
    Throughout this paper, we investigate how different choices of the input physics affect the synthetic stellar populations of the SMC and LMC. To do so, we created the following sets of stellar evolution models:
    \begin{itemize}
        \item {Model Eject:} This is our standard model. It uses the input physics as described in the previous sections, including the modeling of Eddington-limit induced mass ejections. These models assume a constant star-formation history (SFH). We created an additional synthetic population for the SMC called SMC-SFH, where we assume that star formation ended $3.5\,\mathrm{Myr}$ ago to match the observed population better.
        \item {Model no-Eject:} This model has the same setup as detailed in Sect.~\ref{sec:setup}, but without the Eddington-limit induced mass ejections.     
        \item {Model Binary:} This model is a combination of our Model Eject with the LMC binary grids from \citet{mar2:16} and \citet{pau1:22}. We provide more details in Sect.~\ref{sec:binary_pop}.
        \item  {Model RSG:} Within this model, we switch to the RSG mass-loss recipe of \citet{nie1:90} that is frequently employed in stellar evolution models. Note that these mass-loss rates are weaker than those from \citep{yan1:23}. We still use our Eddington-limit induced mass ejection formalism here.
        \item {Model sc10:} This model has the same setup as in Model Eject, but we changed the efficiency of semiconvective mixing to $\alpha_\mathrm{sc}=10$.
    \end{itemize}
    A discussion of including or excluding our Eddington-limit induced mass ejections as well as the role of binaries are provided in Sects.~\ref{sec:nolbv} and \ref{sec:binary}, while the impact of choosing a different physical setup is discussed in Appendix~\ref{sec:add_changes}.
    
    \subsection{Population synthesis}

    \subsubsection{Single stars}
    To test our physical setup, we create synthetic populations of single stars per initial metallicity and per model variation, using the different single-star grid of detailed MESA models (see Sects.~\ref{sec:setup} and \ref{sec:models}). The resulting synthetic populations are compared to the observed populations of the LMC and SMC. 
    For our each of our single-star populations, each of the 33 detailed single-star models of a grid is assigned a statistical weight based on its formation probability and the duration of each time step. The probability of forming a star with a specific mass is determined by the initial mass function (IMF; \citealt{sal1:55}). In this work, we adopt a \citet{kro1:01} IMF and, for simplicity, a constant star formation rate (SFR) of $\mathrm{SFR}_\mathrm{LMC}=0.1\,\msunpyr$ for the LMC \citep[corrected for a Kroupa IMF]{har1:09} and ${\mathrm{SFR}_\mathrm{SMC}=0.05\,\msunpyr}$ for the SMC \citep{rub1:18}. The time spent in a specific evolutionary phase is determined by the detailed evolutionary models and does not rely on any external assumptions. More details on how the weights for individual models are calculated can be found in Appendix~\ref{app:single_star}.

    \subsubsection{Binary stars}
    \label{sec:binary_pop}
    
    Most massive stars are born in binary or higher-order multiple systems \citep{san2:12, san1:13, san1:14, san1:25,dun1:15, fro1:25,vil1:25}. If they have sufficiently close separations, the components interact during their lifetimes. Calculating a large grid of binary models spanning the full initial period and initial mass ratio parameter space is beyond the scope of this paper. However, to test the impact of binarity, we combine our single-star grids with already calculated MESA binary evolutionary grids at LMC metallicity. For the initial primary mass range from $M_\mathrm{1,\,ini}\approx10\,\msun\text{\,--\,}25\,\msun$, we employ the binary evolutionary grid presented in \citet{mar2:16}, and for the range $M_\mathrm{1,\,ini}\approx28\,\msun\text{\,--\,}89\,\msun$, we use the grid from \citet{pau1:22}, which reproduces the observed luminosity distribution of the LMC WR population. These grids have the same mass spacing ($\Delta\log(M_\mathrm{ini}/\msun)=0.05$) as our single-star grids. The models from \citet{mar2:16} span an initial period range from $\log(P_\mathrm{orb,\,ini}/\mathrm{d})=0.15\,\text{--}\,3.5$, while the more massive star models from \citet{pau1:22} that can evolve to larger stellar radii span initial periods from $\log(P_\mathrm{orb,\,ini}/\mathrm{d})=0.15\,\text{--}\,4.0$. Both grids cover the full initial period space where binary interactions occur. We use a period spacing of $\Delta\log(P_\mathrm{orb,\,ini}/\mathrm{d})=0.05$. Furthermore, both grids cover initial mass ratios in the range from $q_\mathrm{ini}=0.25\,\text{--}\,0.95$, with a spacing of $\Delta q_\mathrm{ini}=0.05$. In total, these binary grids encompass about 18\,000 detailed MESA models. To model mass-transfer, both binary model grids employ the ``contact'' scheme of MESA, allowing, as the name suggests, for contact phases. Furthermore, it is assumed that the accretion of mass is rotationally limited, leading to quasi-conservative mass-transfer in short-period systems and non-conservative mass-transfer in wide-period systems. Note that these grids use some different physical assumptions compared to our single-star models, such as mass-loss rates, and thus, detailed comparisons should be made with caution.

    To create the binary population, we assume a binary fraction of $100\%$ and initial orbital periods ranging from ${P_\mathrm{orb,\,ini}=\SI{1.4}{d}\text{\,--\,}\SI{100}{yr}}$ ($\log(P_\mathrm{orb,\,ini}/\mathrm{d})=0.15\,\text{--}\,4.55$. From the binary grids of \citet{mar2:16} and \citet{pau1:22}, we know that binary interactions, such as mass exchange, occur only in systems with $P_\mathrm{orb,\,ini}\lesssim\SI{3000}{d}$, while stars in systems with longer periods evolve effectively as single stars. Following our assumed initial orbital period distribution, this translates to an initial interacting binary fraction of $\approx70\%$, which aligns with observations of massive stars \citep{san2:12,san1:25,vil1:25}. For all models that do not undergo binary interactions, we use our single-star models, meaning that for the synthetic binary population 40\,000 different model combinations were used. Since our Eddington-limit induced mass ejections set an upper limit on a star's radius, we check if a primary star's model at a given initial mass and initial orbital period exceeds the radius of the corresponding single star model. If the primary’s radius grows larger, we assume that the star will be stripped by the Eddington-limit induced mass ejections rather than by mass transfer, and treat the evolution of the system as two independent single stars. For binary systems with initial masses beyond the range covered by the binary models, we assume that if the stars fit within the orbit upon their formation, their evolution can be approximated by single-star models. At these high masses, binary interactions have a less important role, as the maximum radius of a star with $M_\mathrm{ini}=100\,\msun$ is limited to $R_\mathrm{max}\approx58\,\rsun$ due to the Eddington-limit induced mass ejections, which prevents most mass-transfer phases.
    
    As with the single-star population, each binary model is assigned a statistical weight. The initial primary masses are assumed to follow the \citet{kro1:01} IMF, and we adopt the same SFR as for the single star population of the LMC. For simplicity, we assume a flat mass-ratio distribution \citep[i.e., consistent with the distribution reported by][]{san2:12}. The initial orbital period distribution is proportional to $(\log P_\mathrm{orb,\,ini})^{-0.55}$, following \citet{san2:12}. More details on the calculation can be found in Appendix~\ref{app:binary_star}.    

    The binary grids used here assume several criteria for mass-transfer stability, such as L2 overflow during the contact phase, inverse mass-transfer, a limit on the maximum accretion onto the companion based on their ability to expel non-accreted material from the system, and when the mass-transferrate exceeds a high limit, of $0.1\msunpyr$ \citep[for more details, see][]{mar2:16, pau1:22}. If mass transfer is flagged as unstable, we assume that the two stars will merge. If both stars are still on the main sequence (MS+MS merger), we calculate the mass of the merger as ${M_\mathrm{merger,\,ini}=(1-\Phi)(M_\mathrm{1,\,\mathrm{current}}+M_\mathrm{2,\,\mathrm{current}})}$, where the fraction of mass lost during the merger is $\Phi = 0.3q/(1+q)^2$ \citep{gle1:13, wan1:22}. The merger product is then approximated by the closest model with the respective initial mass from our single-star grid. If one of the stars has already exhausted H in its core, we assume that the merger product should also have a helium core and an H-rich envelope (He-MS merger). To estimate the mass of the merger in such cases, we assume that during the potential common envelope phase, half of the envelope of the core-He burning star is lost. The mass of the merger is then $M_\mathrm{ini,\,merger}=(3M_1-M_\mathrm{1,\,he-core})/2+M_2$, and we select the nearest single-star model from our grid after core hydrogen depletion. This is a simplification, and in reality, these mergers should have overmassive H-rich envelopes that would end up as blue supergiants.

    To assess the number of stars in our binary population and compare them with the observed population, we only count the primary stars, the secondaries where the primary star has already died, and the mergers. Secondaries with a primary companion are assumed to be outshone by the companion and should also not be included in the number of observed stars.

    \subsection{Definitions of distinct stellar groups}
    \label{sec:phases}

\begin{figure}[t]
        \centering
        \begin{tikzpicture}[scale=1]
            \node[anchor=south] at (-6.7, 0) {\includegraphics[trim={0.1cm 0.6cm 20cm 1.2cm},clip,height=0.31\textheight]{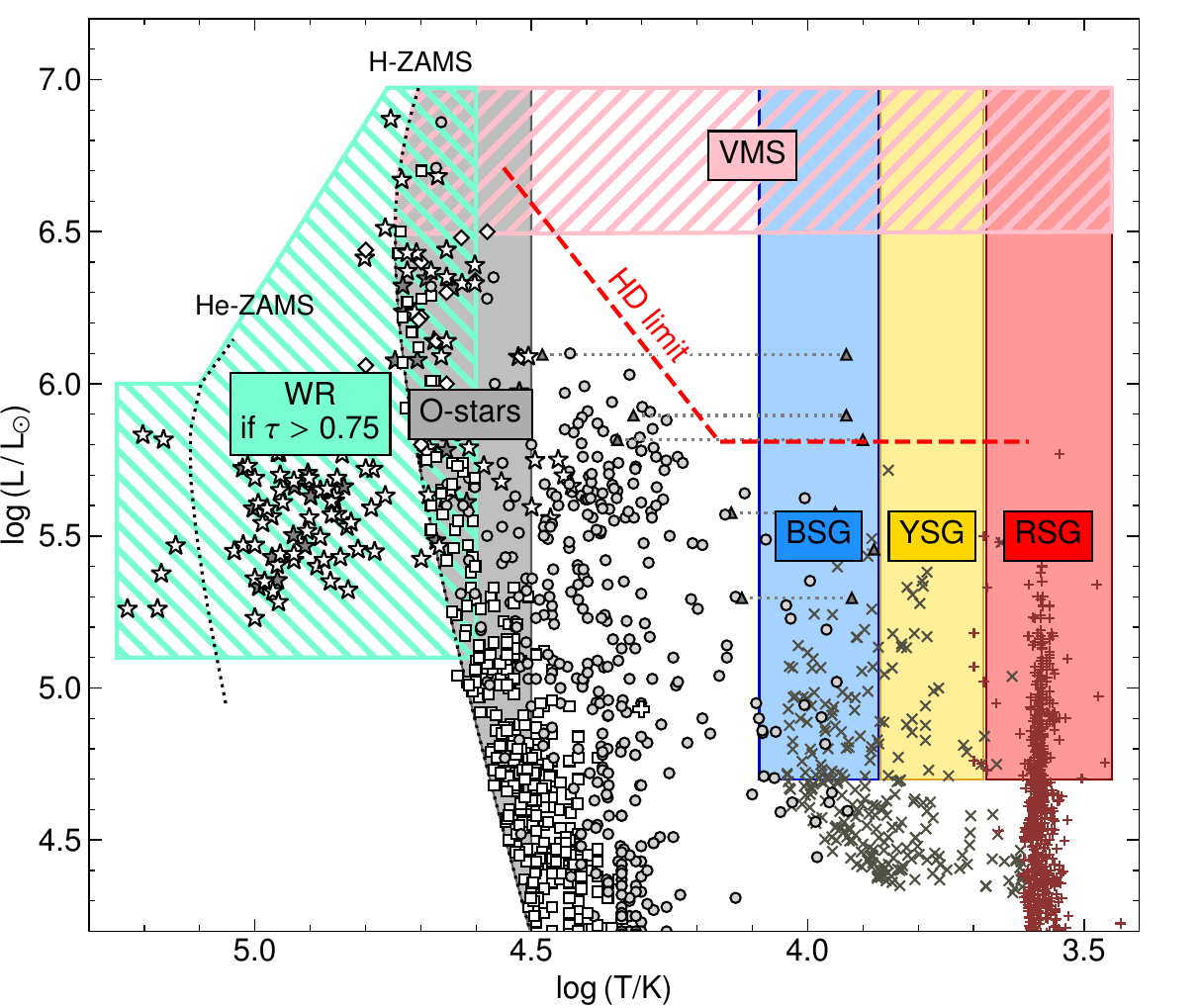}};
        \end{tikzpicture}
        \caption{Sketch of the regions in the HRD where the different stellar groups are located. For reference, in the background, the stellar content of the LMC is shown. The symbols have the same meaning as in Fig.~\ref{fig:hrd_inflation}.}
        \label{fig:groups}
    \end{figure}

    For a consistent comparison of observations to the synthetic populations, we apply selection criteria to both populations to differentiate between the various evolutionary phases. The evolutionary phases considered in this work include WR stars, along with their multiple subtypes, as well as O-type stars, very massive stars (VMSs), and supergiants that are blue (BSG), yellow (YSG), and red (RSG). Figure~\ref{fig:groups} provides an overview of the regions occupied by the different stellar groups. A detailed explanation on how stars are assigned to each group is provided in Appendix~\ref{app:groups}.
    
    \section{Results}
    \label{sec:result}

    In this section, the properties and characteristics of the synthetic population using our new Eddington-limit induced mass ejection formalism, namely Model Eject, are described. A summary of the number of stars during the distinct evolutionary phases of the observed and all synthetic populations is provided in Table~\ref{tab:number_of_stars}. Figure~\ref{fig:hrd_lbv} shows HRDs comparing the observed populations of the LMC and SMC with the synthetic populations generated using our Model Eject.

    \subsection{The LMC population}

    For the LMC, the synthetic single-star population contains about 2650 O-type stars, in agreement with the estimated 2780 O-type stars in the LMC (see footnote in Table~\ref{tab:number_of_stars}). The predicted 235 RSGs are also in good agreement with the 262 observed RSGs. Figure~\ref{fig:rsg_lbv} compares the predicted and empirical luminosity distributions of BSGs, YSGs, and RSGs in both the LMC and SMC. Not only is the total number of RSGs well matched, but their luminosity distribution is also closely reproduced. 

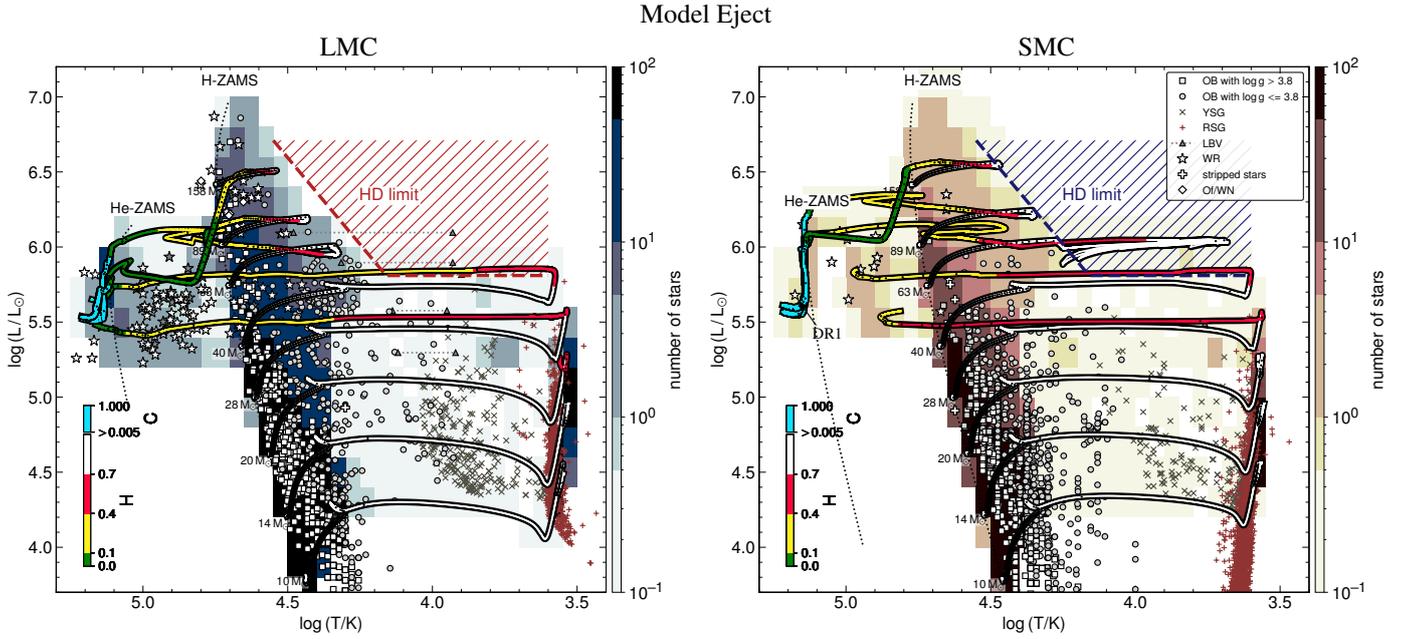
\begin{figure*}[tbhp]
        \centering
        \begin{tikzpicture}
            \node[anchor=south] at (0, 0) {\includegraphics[trim={0cm 0.75cm 0cm 1.2cm},clip,height=0.31\textheight]{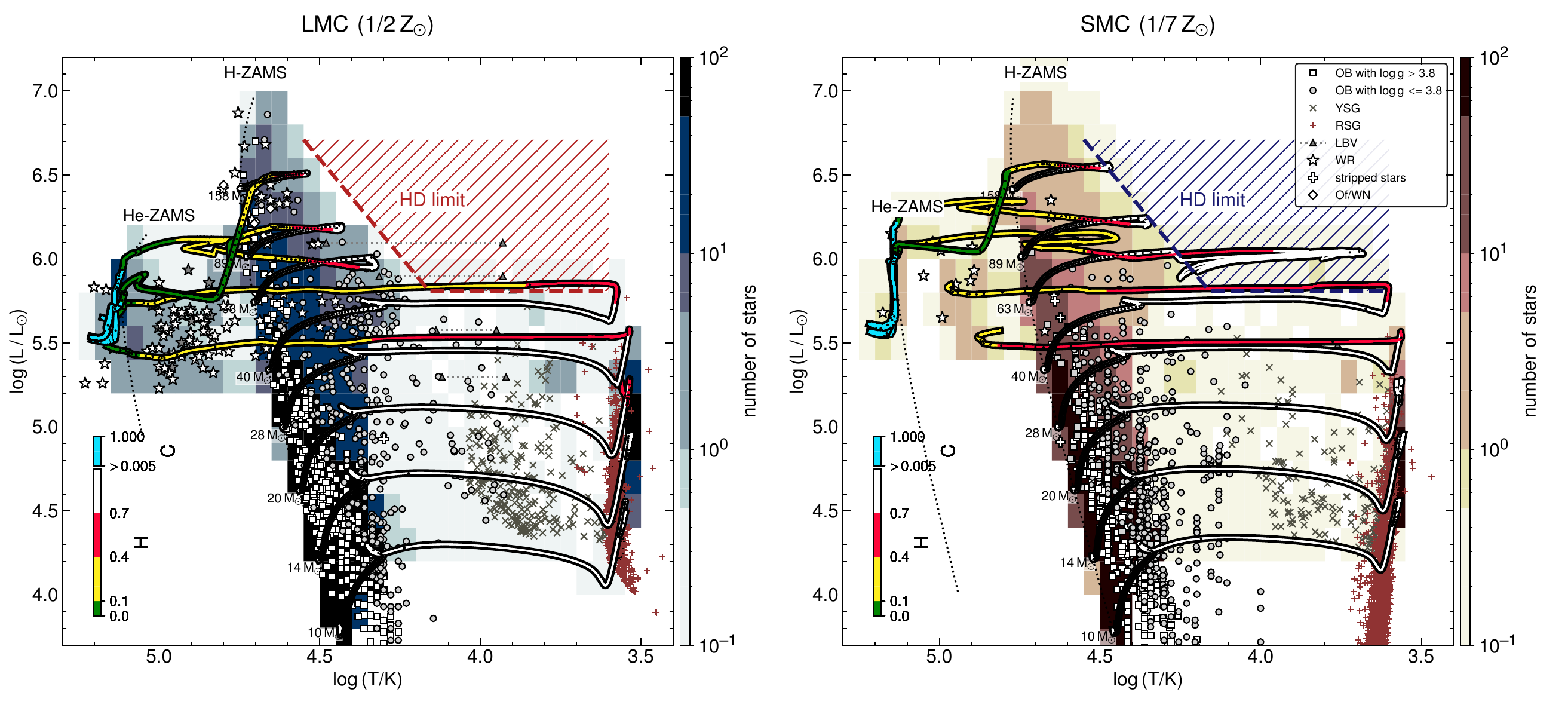}};
            \node at (0, 8.3) {Model Eject}; 
            \node at (-4.7, 7.9) {LMC};
            \node at (4.48, 7.9) {SMC};
            \node at (1.6, 4.1) {\fontsize{6}{14}\selectfont DR1};
            \draw[-latex] (1.45,4.2) -- (1.2,4.55);
        \end{tikzpicture}
        \caption{Hertzsprung-Russell diagrams of the massive star population (symbols) in the LMC (left) and SMC (right) compared to the synthetic population of the Model Eject (contours). A selection of the same stellar evolution models is shown as solid lines, color coded by their surface abundances. Small black dots on the evolution tracks mark equidistant timesteps of $\Delta t = \SI{30000}{yr}$. The hatched region indicates the empirical HD limit \citep{hum1:79}. The different symbols mark different stellar groups and have the same sources as mentioned in Fig.~\ref{fig:hrd_inflation}. In the SMC's population, we added the WO star DR1 from IC\,1613 \citep{tra1:13}, a galaxy that has an SMC-like metallicity and only this WR star.}
        \label{fig:hrd_lbv}
    \end{figure*}

\begin{figure}[tbhp]
        \centering
        \begin{tikzpicture}[scale=0.9]
            \node[anchor=south] at (0, 0) {\includegraphics[trim={0.1cm 0.3cm 0cm 0cm},clip,height=0.22\textheight]{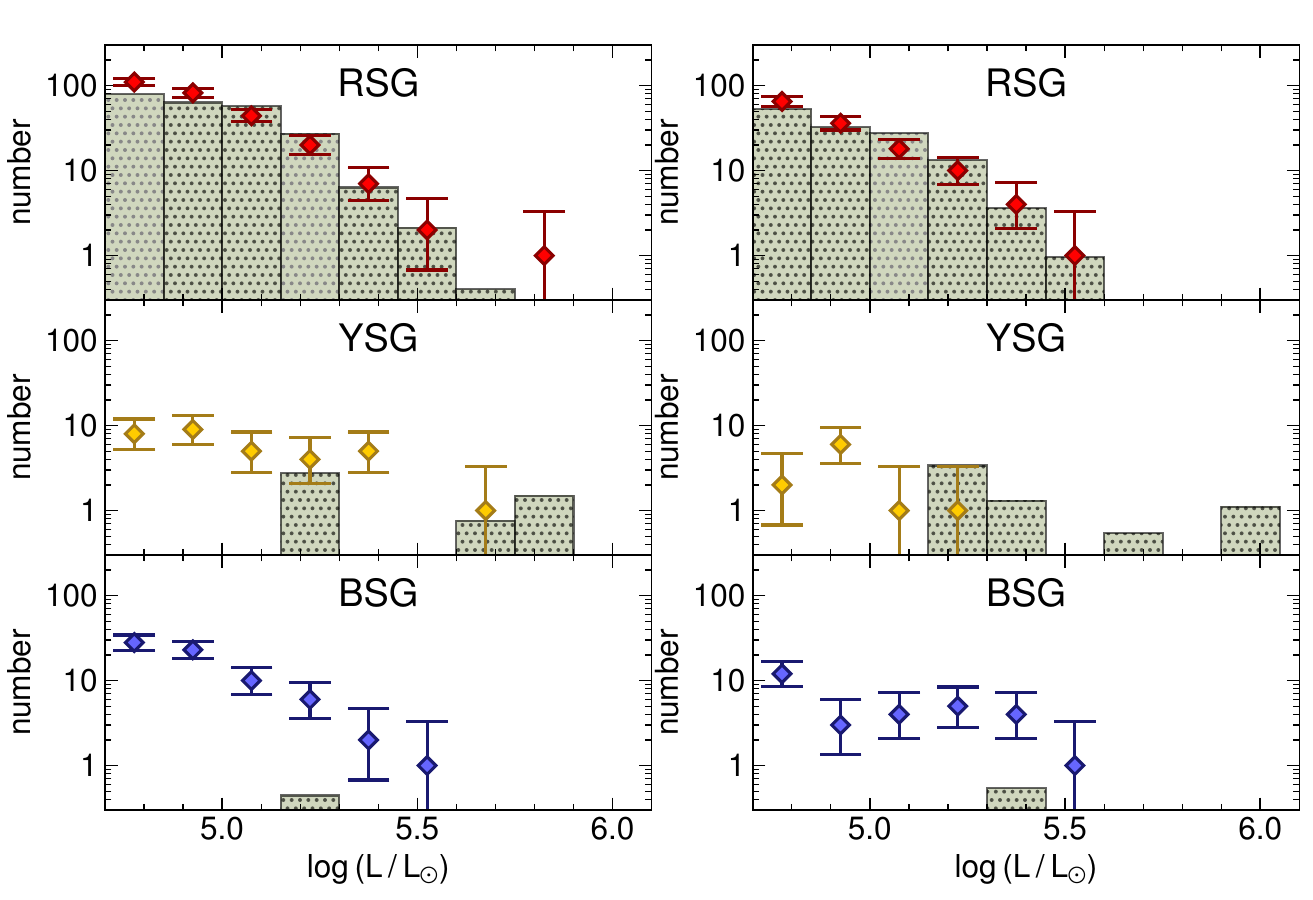}};
            \node at (0.2, 6.6) {Model Eject}; 
            \node at (-1.95, 6.3) {LMC};
            \node at (2.55, 6.3) {SMC};
        \end{tikzpicture}
        \caption{Empirical luminosity distributions for the different cool supergiants (diamonds) in the LMC (left) and SMC (right) compared to the predictions from the Model Eject (hatched bins). The error margins represent the upper and lower bounds of the 84.13\% confidence interval of a Poisson distribution, accounting for small numbers \citep{isr1:68}. The sources of the observed stars are the same as in Fig.~\ref{fig:hrd_inflation}.}
        \label{fig:rsg_lbv}
    \end{figure}

    From the HRD shown in Fig.~\ref{fig:hrd_lbv}, one can see that stars with $M_\mathrm{ini}\sim25\,\msun\,\text{--}\,50\,\msun$ lose their H-rich envelopes during the RSG phase and evolve into WR stars. In these models, after the RSG phase, when most of the H-rich envelope has been removed and the envelope switches from convective to radiative, the inflation criterion for the mass ejections is triggered (see the $28\,\msun$ and $40\,\msun$ tracks shown in Figs.~\ref{fig:hrd_phases}). This outburst typically removes less than one solar mass, and occurs in a region close to the S\,Dor instability strip where the faintest LBVs are observed. \citet{mar1:25} examined the spatial distribution of LBVs in the LMC and report that the faint LBVs in the LMC are associated with evolved stars, and that they are most likely core-He burning, which is in agreement with our model predictions.

    \begin{table*}[thb]
            \centering
            \caption{Summary of the number of stars in a specific stellar group for the different stellar populations.}
            \small
            \begin{tabular}{l|cc|ccc|cc|cc|cc|c}\hline \hline \rule{0cm}{2.8ex}%
                \rule{0cm}{2.2ex} Group & \multicolumn{2}{c|}{observed$^\dagger$}  & \multicolumn{3}{c|}{Model Eject}         & \multicolumn{2}{c|}{Model no-Eject} & \multicolumn{2}{c|}{Model RSG} & \multicolumn{2}{c|}{Model sc10} & \multicolumn{1}{c}{Model Binary}  \\
                \rule{0cm}{2.2ex}       & LMC            & SMC           & LMC            & SMC           & SMC-SFH         & LMC             & SMC             & LMC            & SMC           & LMC             & SMC           & LMC                              \\
                \hline \rule{0cm}{2.4ex}%
                    VMS                 & 13             & 0             & 27.7           & 16.5          & 0               & 44.5            & 22.8            & 29.2           & 19.5          & 28.9            & 16.9          & 18.7 \\
                    WR                  & 122$^{(b)}$    & 12            & 132.5          & 28.5          & 12.5            & 25.9            & 0               & 87.4           & 19.5          & 79.7            & 23.1          & 120.2 \\
                    H-rich WN           & 5              & 2             & 8.0            & 1.9           & 0.1             & 0               & 0               & 8.0            & 1.7           & 7.8             & 1.8           & 6.4 \\
                    H-poor WN           & 50             & 9             & 69.0           & 14.5          & 7.8             & 24.4            & 0               & 44.4           & 9.6           & 35.0            & 9.5           & 50.9 \\
                    H-free WN           & 41             & 0             & 41.5           & 4.8           & 1.1             & 1.5             & 0               & 23.3           & 4.6           & 22.8            & 5.0           & 46.6 \\
                    WC/WO               & 26$^{(c)}$     & 1             & 14.4           & 7.3           & 3.5             & 0               & 0               & 11.8           & 3.6           & 14.2            & 6.7           & 16.4 \\
                    O-stars             & $\sim$2780$^{(d)}$& $\sim$500$^{(e)}$& 2653     & 1789          & 887             & 2548            & 1756            & 2612           & 1764          & 2601            & 1765          & 3198 \\
                    BSG                 & 70             & 28            & 1.4            & 1.1           & 1.0             & 5.8             & 2.3             & 1.0            & 0.7           & 9.7             & 18.8          & 3.7\\
                    YSG                 & 32             & 10            & 5.5            & 6.7           & 6.7             & 29.2            & 22.6            & 7.1            & 7.9           & 57.8            & 13.9          & 8.05\\
                    RSG                 & 262            & 134           & 235.8          & 130.9         & 130.9           & 235.3           & 133.3           & 280.0          & 154.8         & 199.0           & 80.6          & 270.5\\
                \hline
            \end{tabular}
            \rule{0cm}{2.8ex}%
            \begin{minipage}{0.95\linewidth}
                \ignorespaces 
                 $^\dagger$ If not mentioned otherwise, the works used to calculate the numbers of observed stars are the same as the ones quoted in Fig.~\ref{fig:hrd_inflation}. $^{(b)}$ Stars with spectral type Of/WN were excluded from this number, as these are not covered by our $\tau$-criterion (see Appendix~\ref{app:groups}). $^{(c)}$ Only 7 of the WC/WO stars have been analyzed with non-local thermodynamic equilibrium stellar atmosphere models, but 26 are confirmed \citep{bre1:81,tes1:93,bar1:82,neu1:12}. $^{(d)}$ \citet{dor1:13} investigated the completeness of the O-star content in the 30\,Dor region of the LMC and reported the presence of about $N_\mathrm{O,\,30Dor}=570$ O stars. Furthermore, we know  that 30\,Dor hosts $N_\mathrm{WR,\,30Dor}=25$ WN/WC/WO stars \citep[][and references therein]{dor1:13}. Assuming that WR stars (excluding the extremely young Of/WN stars) are tracers of massive star populations, we infer that the number of O-stars in the LMC can be approximated as $N_\mathrm{O,\,LMC}=N_\mathrm{WR,\,LMC}/N_\mathrm{WR,\,30Dor}\cdot N_\mathrm{O,\,30Dor}$. The number quoted here corresponds to the average of this range. $^{(e)}$ The BLOeM sample, i.e., the largest spectroscopic sample of massive stars in the SMC, contains 159 O-stars and has an estimated completeness of around $\sim35\%$ compared to Gaia DR3. This implies that there are $\sim500$ O-type stars in the SMC \citep{she1:24}. Note that this number might be larger since Gaia cannot resolve the dense cores of young clusters.
            \end{minipage}
        \label{tab:number_of_stars}
    \end{table*}
    
    Models with $M_\mathrm{ini}\gtrsim50\,\msun$ undergo an ejection episode already during core-H burning, preventing them from crossing the HD limit. Upon entering this phase, the models fall out of thermal equilibrium and reach mass-loss rates of $\log(\dot{M}_\mathrm{eject}/(\msunpyr))\approx-2$ to $-3$. These high rates quickly stop the stellar models from inflating, such that they can regain thermal equilibrium. Following the initial rapid mass ejection, the models remain close to their inflation limit ($R/R\mathrm{core}=1.9$) with lower but still substantial rates of $\log(\dot{M}/(\msunpyr))\approx-4.5$, comparable to luminous WNh stars located to the right of the ZAMS. During the mass ejection episode, large fractions of the H-rich envelope are shed. Depending on initial mass, $\approx20\,\msun\text{\,--\,}300\,\msun$ can be lost.

    After core-H depletion, the models expand again (regardless of whether or not MESA's superadiabatic reduction method has been used), triggering a second mass ejection episode. This leads to an intense but short-lived mass-loss episode in which the remaining H-rich envelope (less than a few solar masses) is removed, producing hot H-free WN stars that soon evolve into WC stars. While this outburst phase may be too brief to be observed directly, it should leave behind a nebula around the WR star. Interestingly, three WR stars in the LMC are known to have nebulae \citep{sto1:11}: one is a single hot early-type WN, one a single WC, and one a WC in a binary. Our model predictions agree with the WR temperatures and luminosities. However, there is no proof that these stars have gone through an Eddington-limit induced mass ejections, as the material could also be leftovers from a RSG or common envelope phase.
    
    \begin{figure}[tbhp]
        \centering
        \begin{tikzpicture}[scale=1]
            \node[anchor=south] at (0, 0) {\includegraphics[trim={0.1cm 0.75cm 0cm 0cm},clip,height=0.32\textheight]{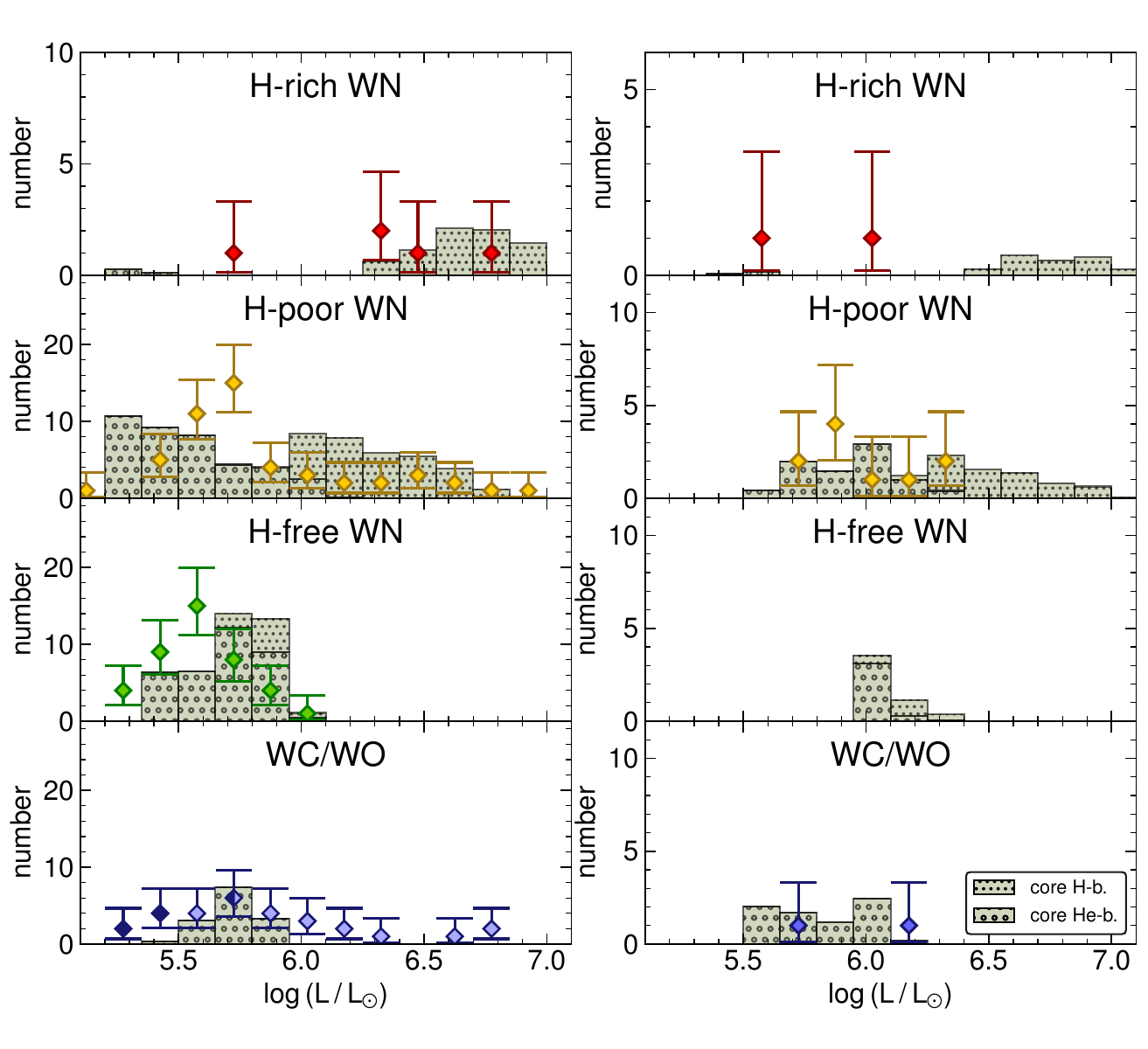}};
            \node at (0.2, 8.2) {Model Eject}; 
            \node at (-1.95, 7.9) {LMC};
            \node at (2.55, 7.9) {SMC};
            \node at (1.5, 1.7) {\fontsize{7}{14}\selectfont DR1};
            \draw[-latex] (1.55,1.5) -- (1.65,1.2);
        \end{tikzpicture}
        \caption{Empirical luminosity distributions for the different WR subtypes (diamonds) in the LMC (left) and SMC (right) compared to the predictions from the Model Eject population (hatched bins). The error margins mark the 84\% confidence intervals of a Poissonian distribution accounting for small numbers \citep{isr1:68}. In the synthetic population, we differentiate between core H-burning (light dotted bins) and core He-burning stars (bold dotted bins). For definitions of the individual WR subtypes, refer to Appendix~\ref{sec:add_changes}. The Of/WN stars are not included in the H-rich WR subtype, as they do not meet the criteria used in our models to identify the WR phase. For the LMC's WC/WO population, not all stars have been studied in detail using stellar atmosphere models, and for a large set of the WCs, we had to estimate their luminosity from their V-band photometry \citep[see Section 4 of ][]{pau1:22}. Bins containing only luminosities derived from a detailed spectral analysis are marked by dark blue diamonds, those only containing luminosities derived from the V-band photometry are marked by light blue diamonds, and those containing both are marked by half dark and half light blue diamonds. Since some of these stars might be binaries, the luminosities estimated from the V-band should be treated with caution and considered as upper limits. In the SMC's WC/WO population, we added the WO star DR1 from IC\,1613 \citep{tra1:13}.}
        \label{fig:wr_lbv}
    \end{figure}
    
    Deficiencies of our single-star population are the lack of OB supergiants near the terminal age main-sequence as well as the BSG and YSG populations when compared to the observed ones. \citet{san1:13} reported a lack of binaries in the O9.7\,I, suggesting that they could be post-interaction binaries or merger products. Alternatively, the BSG might be explainable when using different assumptions on mixing \citep{gil1:21}. We discuss this in more detail in Sect.~\ref{sec:discuss} and Appendix~\ref{sec:add_changes}. At the high-mass end, the predicted number of 28 VMSs differs by a factor of two from the observed value of 13. At such high masses, the population is very sensitive to the assumed form of the IMF and SFH. Small changes in these assumptions can easily account for the difference of the factor of two.

    The synthetic single-star population of the LMC contains about 132 WR stars, which is close to the observed number of 122 WRs. The different WR subtypes serve as an excellent probe to test the stellar physics used in models, as they are sensitive to the assumptions made about mass-loss and mixing. Figure~\ref{fig:wr_lbv} shows the luminosity distributions of the WR subtypes in the LMC and SMC. Our population covers roughly the luminosity distribution of the observed WR stars and can explain the so far unexplained faintest single WR stars \citep[][and references therein]{she1:20}. 
    There are only small discrepancies: a few too many H-poor WN stars around $\log(L/\lsun)\approx6.1$, which originate from the core hydrogen-burning WN population after the Eddington-limit induced mass ejections, and too few bright H-free WNs around $\log(L/\lsun)\approx5.8$, which originate from stars experiencing mass ejections. Both of these features, as well as the overabundance of VMSs in the synthetic population compared to observations, might suggest that our assumptions on the IMF or SFH are too simplistic and thus lead to such artifacts. The low number of predicted fainter H-free WN stars in the LMC population may be related to our choice of when to transition from the hot star mass-loss recipe to the WR mass-loss recipe. However, shifting this transition to lower $\Gamma_\text{\!e}$ values would lead to an overproduction of WN stars in the SMC, which is inconsistent with observations. Another crucial factor is the impact of binary interactions, which can lead to a more efficient removal of the H-rich envelope, making it easier for the WN stars to become H-free. \citet{pau1:22} report that binary interactions could be the primary mechanism behind the WN and WC populations in the LMC. We discuss this further in Sect.~\ref{sec:binary}. For the WC/WO population in the LMC, our models predict that only stars that undergo a mass ejection episode evolve into WC/WO-type stars. Note that the brightest WC/WO stars in the LMC are most likely binaries, leading to too high luminosities when applying the bolometric correction \citep{bar1:01,bar2:01,pau1:22}.

    \subsection{The SMC population}

    \begin{figure*}[tbhp]
        \centering
        \begin{tikzpicture}[scale=1]
            \node[anchor=south] at (-6.7, 0) {\includegraphics[trim={22.4cm 0.75cm 0cm 1.2cm},clip,height=0.31\textheight]{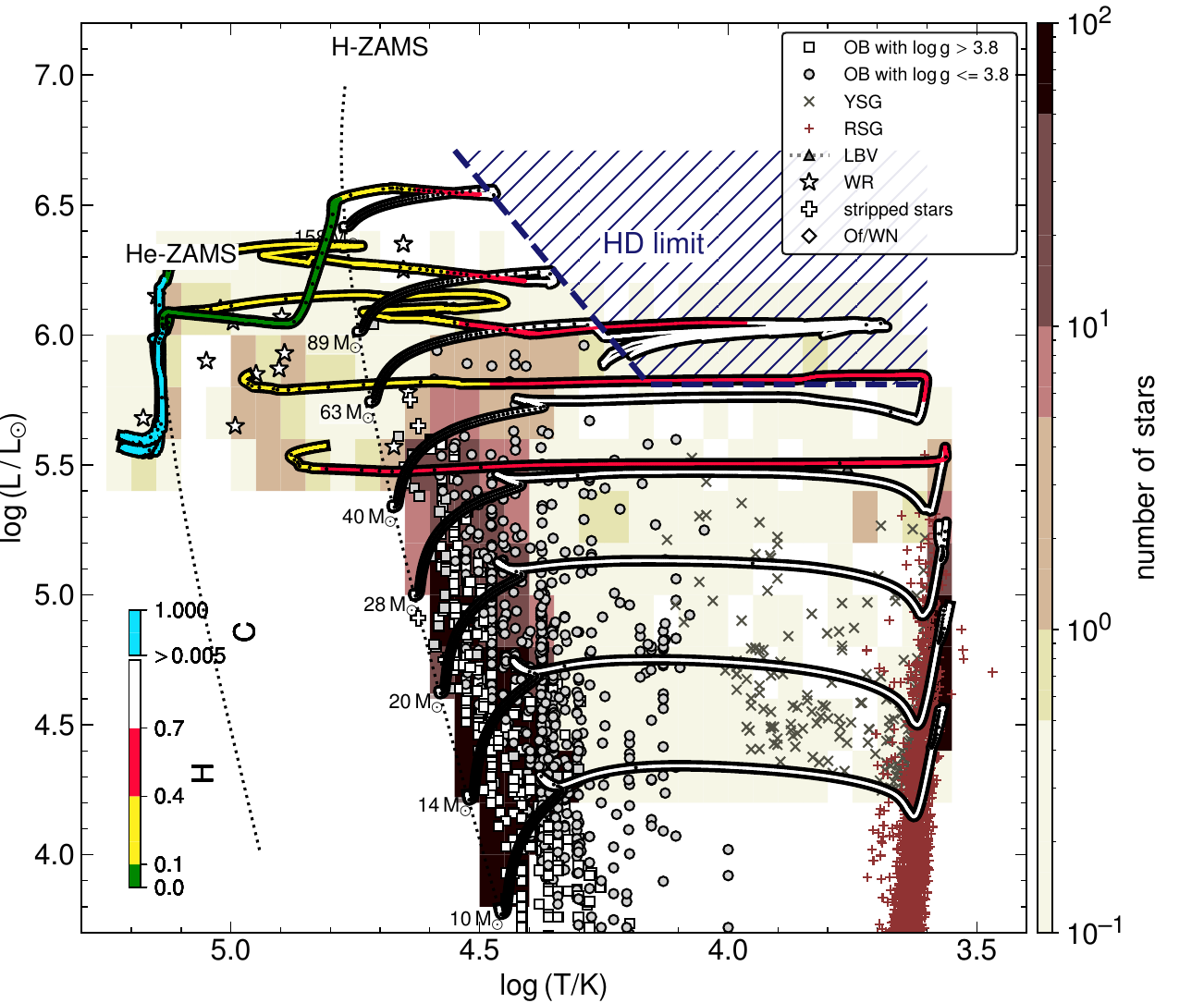}};
            \node at (4.48-11.2, 7.9) {SMC-SFH};
            \node at (1.6-11.4, 4.1) {\fontsize{6}{14}\selectfont DR1};
            \draw[-latex] (1.45-11.4,4.2) -- (1.2-11.4,4.55);
            \node[anchor=south] at (0, 0) {\includegraphics[trim={11.2cm 0.75cm 0cm 0cm},clip,height=0.32\textheight]{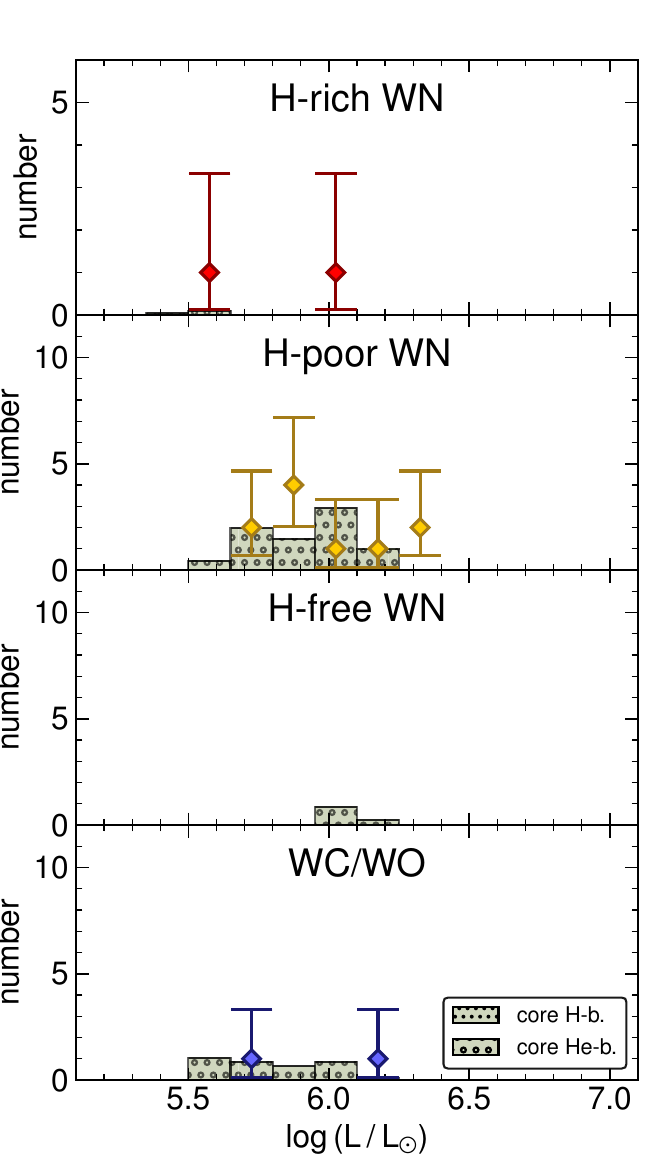}};
            \node at (-4, 8.2) {Model Eject}; 
            \node at (0.225, 7.9) {SMC-SFH};
            \node at (1.5-2.55+0.225, 1.7) {\fontsize{7}{14}\selectfont DR1};
            \draw[-latex] (1.55-2.55+0.225,1.5) -- (1.65-2.55+0.225,1.2);
        \end{tikzpicture}
        \caption{Hertzsprung-Russell diagram of the SMC population (left) and luminosity distributions of the WR subtypes in the SMC (right) compared to the model predictions of our Model Eject, where star-formation stopped $3.5\,\mathrm{Myr}$ ago. In the SMC's WC/WO population, we added the WO star DR1 from IC\,1613 \citep{tra1:13}. The symbols and colors are the same as in Figs.~\ref{fig:hrd_lbv} and \ref{fig:wr_lbv}.}
        \label{fig:smc-sfh}
    \end{figure*}

    The HRD for SMC metallicity, shown in Fig.~\ref{fig:hrd_lbv}, reveals a somewhat different picture compared to the LMC. The synthetic population predicts about 1800 O-type stars, roughly three times as many as are observed. This discrepancy primarily arises from the absence of stars on the first half of the main sequence for initial masses above $M_\mathrm{ini}\gtrsim25\,\msun$. This phenomenon was studied in more detail by \citet{sch1:21}, who conclude that changes in the SFH or the IMF alone cannot explain the observed discrepancy. They conclude that another process is likely at work, and propose that the absent stars might be hidden in their birth clouds.

    The RSG population, mainly originating from stars with $M_\mathrm{ini}\lesssim25\,\msun$, contains 131 stars in the synthetic population, closely matching the observed 124 RSGs and their luminosity distribution (Fig.~\ref{fig:rsg_lbv}). At SMC metallicity, OB-star winds are weaker, leading to less mass removal from the H-rich envelope during core-H burning. Consequently, more envelope mass must be lost during the RSG phase, raising the minimum initial mass required to form WR stars to $M_\mathrm{ini}\gtrsim28\,\msun$. Interestingly, the luminosity of our faintest WR star aligns with the faintest observed single WR in the SMC \citep{hai1:15, sch1:24}, which previous single-star models could not explain.

    As in the LMC, models with $M_\mathrm{ini}\gtrsim50\,\msun$ experience a mass ejection episode, preventing most stars from exceeding the HD limit. Envelope inflation is $Z$-dependent, causing stars to expand less rapidly at lower metallicity. This can be seen by a few models that surpass the HD limit for less than a few thousand years. In total, one star is predicted beyond the HD limit, consistent with observations. Given the weak $Z$ dependence of the Eddington-limit induced mass ejection prescription, it might still be possible that below SMC metallicity stars can be found beyond the HD limit. However, this number should still be much smaller compared to models that ignore Eddington-limit induced mass ejections (see Sect.~\ref{sec:nolbv}). Furthermore, the synthetic SMC population predicts 17 VMSs, whereas none have been observed. Although we are working within a parameter regime where low-number statistics are relevant, the absence of any observed stars that should be the brightest in the galaxy is concerning.

    Examining the WR subtype luminosity distributions, displayed in Fig.~\ref{fig:wr_lbv}, the synthetic population generally reproduces the observed distributions of the H-rich and H-poor WNs but fails at the H-free WNs and WC/WO stars. In general, it also overpredicts the total number of WR stars by a factor of three, which is mostly due to the prediction of too many bright (massive) stars. From the H-rich WN population, two key points emerge. First, the synthetic population tends to predict too many bright WR stars. Second, our models struggle to reproduce the faint H-rich WN stars. This discrepancy arises from our criterion for defining the WR phase. If one looks at the $M_\mathrm{ini}=28\,\msun$ track in Fig.~\ref{fig:hrd_lbv}, we see that it passes through the region where H-rich WN stars are observed and has the fitting surface H-abundance. For the H-poor WN population, the synthetic model matches the shape of the observed population well, although it is slightly overabundant, especially at the high luminosity end, which is populated by stars going through a mass ejection episode. The additional mass loss efficiently strips the H-rich envelope in the most massive stars, and predicts 4 H-free WNs and 4 WC/WO stars. While the SMC does not contain any H-free WNs, it does host one luminous WO-type star in a binary system. The position of this star matches the upper luminosity range covered by our evolutionary tracks. Interestingly, the evolutionary tracks shown in Fig.~\ref{fig:hrd_lbv} indicate that the WC/WO phase stars extend down to luminosities of about $\log(L/\lsun)\approx5.5$. IC\,1613, a small galaxy with SMC-like metallicity \citep{tau1:07}, hosts one WR star (DR1), a WO-type star with a luminosity of $\log(L/\lsun)=5.68\pm0.10$ \citep{tra1:13}. So far, this star has not been explained by single-star evolution models using mass-loss recipes that match typical observational data. We believe that our Eddington-limit induced mass ejection implementation offers a new possibility to explain this star's evolutionary history.
    
    \subsection{The SMC population with changed SFH}

    The overabundance of O-type stars, VMSs, and WR stars, each arising from stars with $M_\mathrm{ini}\gtrsim28\,\msun$, in the synthetic SMC population, may suggest that our assumption of a constant SFH could be erroneous. To explore this further, we created a new synthetic population, assuming that star formation in the SMC stopped $3.5\,\mathrm{Myr}$ ago. We chose this stopping condition as it matches well the shape of the observed stars in the SMC and is roughly consistent with the age estimates of the bright O-stars in the young cluster NGC~346 \citep{ric1:23}.
    The HRD and WR luminosity distribution for this new population are shown in Fig.~\ref{fig:smc-sfh}.

    \begin{figure*}[tbhp]
        \centering
        \begin{tikzpicture}
            \node[anchor=south] at (0, 0) {\includegraphics[trim={0cm 0.75cm 0cm 1.2cm},clip,height=0.31\textheight]{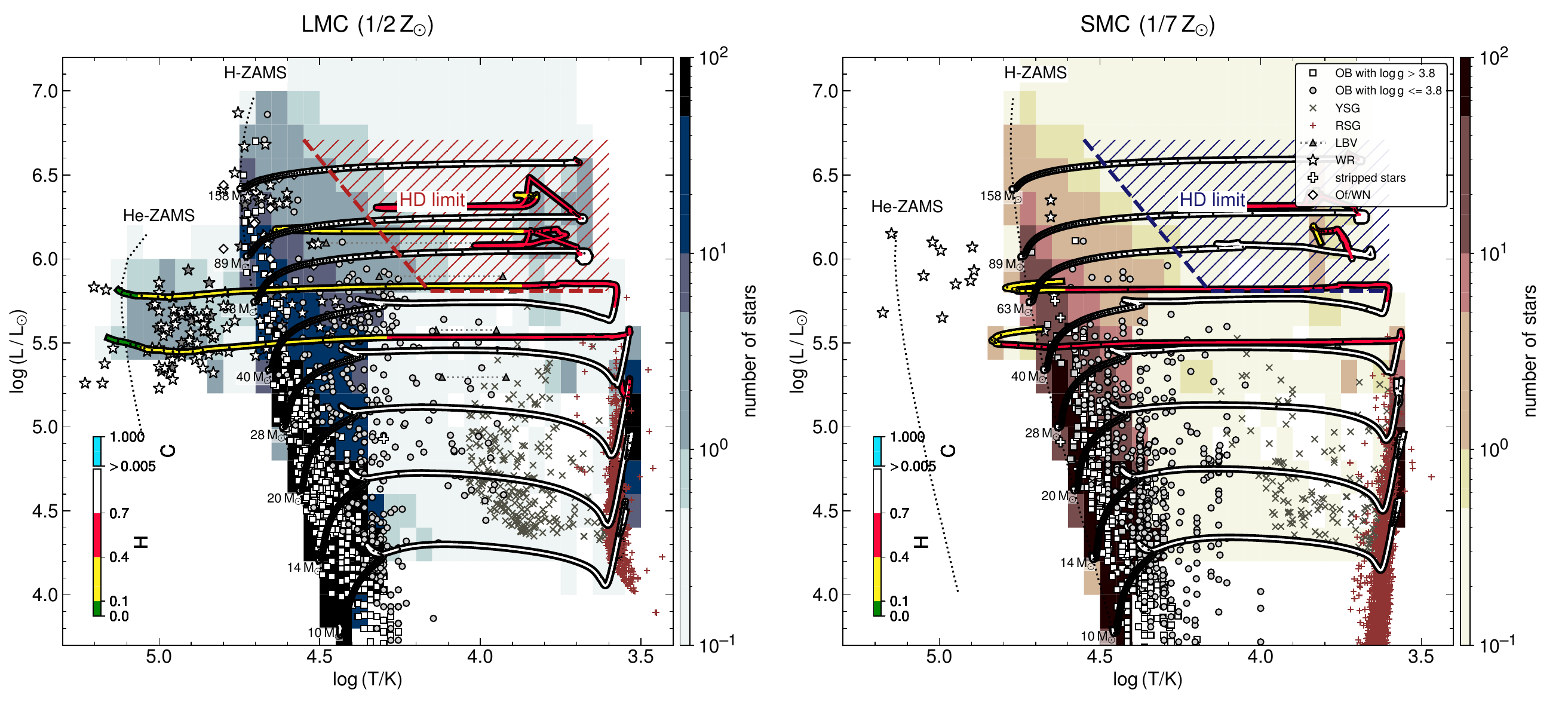}};
            \node at (0, 8.3) {Model no-Eject}; 
            \node at (-4.7, 7.9) {LMC};
            \node at (4.48, 7.9) {SMC};
            \node at (1.6, 4.1) {\fontsize{6}{14}\selectfont DR1};
            \draw[-latex] (1.45,4.2) -- (1.2,4.55);
        \end{tikzpicture}
        \caption{Hertzsprung-Russell diagrams of the massive star population (symbols) in the LMC (left) and SMC (right) compared to the synthetic population of the Model no-Eject (contours). The symbols and colors are the same as in Fig.~\ref{fig:hrd_lbv}.}
        \label{fig:hrd_noLBV}
    \end{figure*}
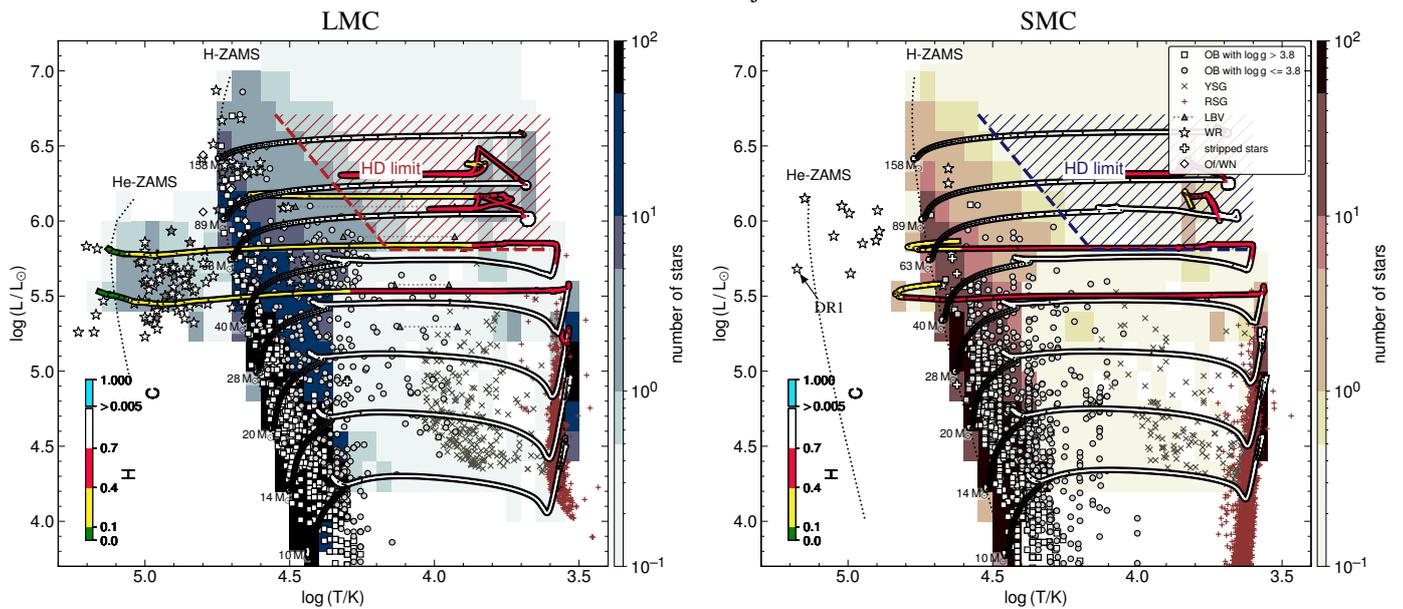

    From the HRD of the synthetic population, illustrated in the left panel of Fig.~\ref{fig:smc-sfh}, it is immediately apparent that the age cut removes all VMSs (their evolutionary lifetime is much shorter than our age cut), bringing the synthetic population into better agreement with observations. Additionally, the region near the ZAMS, for stars with $M_\mathrm{ini}\gtrsim25\,\msun$, is no longer populated, and the contours of the synthetic population now match the observed distribution much more closely. Our synthetic population now contains about 890 O-type stars, which is closer to the estimated number of 500 O-type stars, but still not in full agreement. Upon closer inspection of the HRD shown in Fig.~\ref{fig:smc-sfh}, an absence of O-stars in the spectroscopic sample can be seen near the predicted turn-off point located around $\log(T/\mathrm{kK})\approx4.5$ and $\log(L/\lsun)\approx5.7$. However, this absence is not present in the photometric Gaia sample of \citet[][their figure 5]{sch1:21}, suggesting it might be a coincidence that these stars are missing in our spectroscopic sample. Alternatively, this lack of stars could reflect a binarity effect, where the most luminous stars are binary products, namely post-interaction binaries or mergers. If it were a binary effect, it could also imply that our SFH is still too active and thus might lead to an overestimation of the O-type stars.
    
    \begin{figure}[tbhp]
        \centering
        \begin{tikzpicture}[scale=1]
            \node[anchor=south] at (0, 0) {\includegraphics[trim={0.cm 0.75cm 0cm 4cm},clip,height=0.25\textheight]{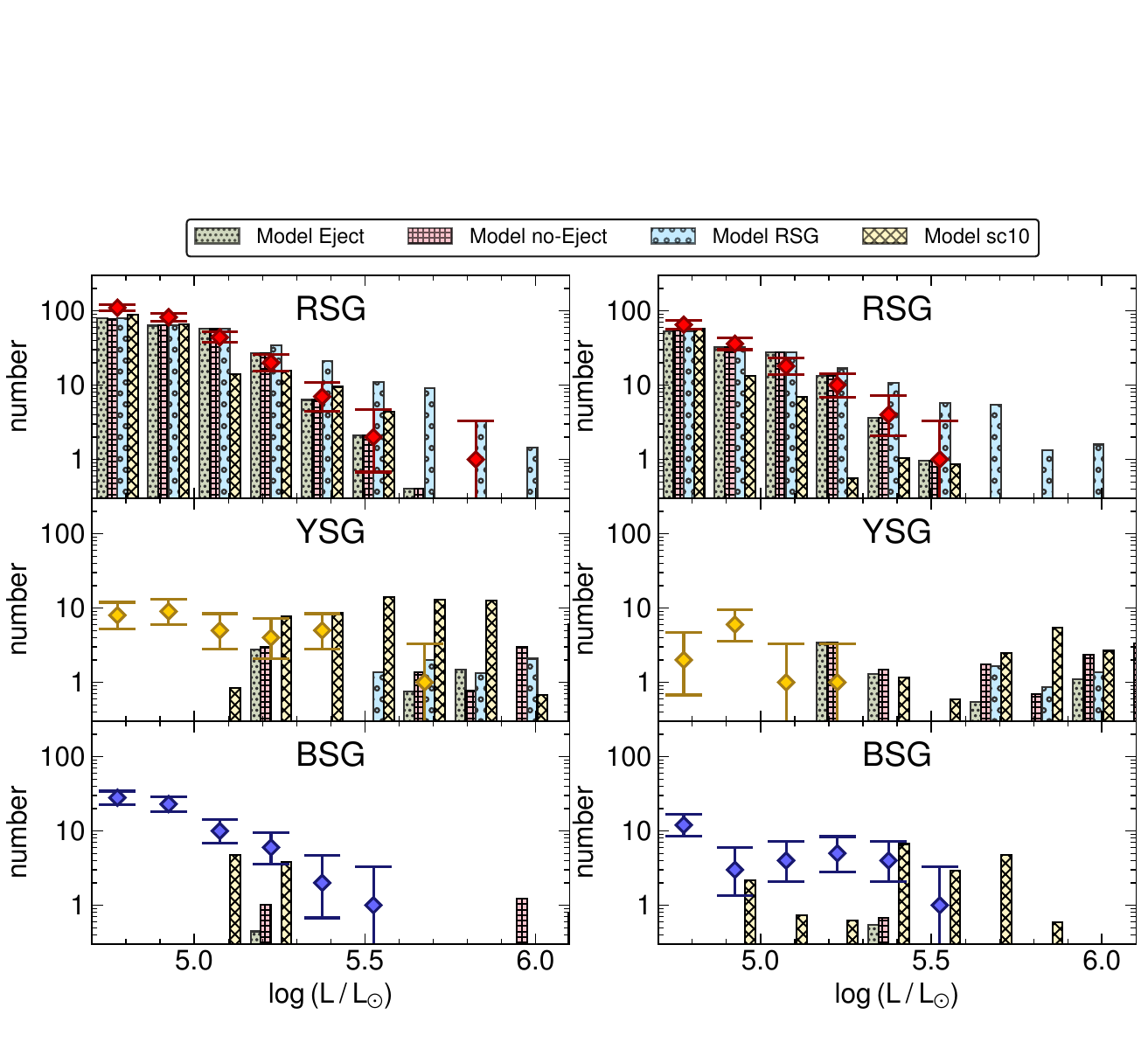}};
            \node at (-1.95, 6.5) {LMC};
            \node at (2.55, 6.5) {SMC};
        \end{tikzpicture}
        \caption{Empirical luminosity distributions for the different cool supergiants (diamonds) in the LMC (left) and SMC (right) compared to the predictions from the models with different physical setups (bins). The sources of the observed stars are the same as in Fig.~\ref{fig:hrd_inflation}.}
        \label{fig:csg_comp}
    \end{figure}
    
    The WR luminosity distribution, shown in the right panel of Fig.~\ref{fig:smc-sfh}, changed substantially compared to the constant-SFR case. The luminous H-rich and H-poor WNs are no longer present in our synthetic population.  With the adjusted SFH, the distribution of H-poor stars is well matched, though we are now missing a few bright WR stars. Interestingly, due to the age cut, only one H-free WN star, which is created as a result of the Eddington-limit induced mass ejections during the evolution of the most massive stars in our model grid, is expected to exist in the SMC. Furthermore, our models still predict three WC/WO stars within the observed luminosity range of the WO star in the SMC. Note that this observed WO star is in a close binary, which might indicate that it is the result of binary interaction rather than Eddington-limit induced mass ejections. However, DR1, the apparently single WO star in IC\,1613, still fits this scenario.

    \section{Discussion}
    \label{sec:discuss}     
    
    \subsection{Neglecting Eddington-limit induced mass ejections}
    \label{sec:nolbv}

    An HRD for a single-star population computed with the same physical setup as before, but without including Eddington-limit induced mass ejections (Model no-Eject), is shown in Fig.~\ref{fig:hrd_noLBV}. In this case, all of our massive star models evolve beyond the HD limit, which conflicts with observations.

    In Fig.~\ref{fig:csg_comp}, we compare the luminosity distribution of the observed cool supergiants to the predictions of different models. Model no-Eject still reproduces the RSG population fairly well. Stars with $M_\mathrm{ini}\lesssim25\,\msun$ (i.e., those not affected by envelope inflation) evolve similarly in both the Eject and no-Eject models, as they never experience a mass ejection episode. Stars with $M_\mathrm{ini}\approx25\,\msun\text{\,--\,}40\,\msun$ still lose most of their H-rich envelope during the RSG phase, but without the mass ejections present in Model Eject (see Fig.~\ref{fig:hrd_phases}). Instead, they must shed these layers through their stellar winds, which is reflected in fewer WR stars.

    For $M_\mathrm{ini}\gtrsim50\,\msun$, the neglect of Eddington-limit induced mass ejections allows the stars to expand beyond the HD limit while still in core hydrogen burning, spending most of their time as YSG. Since we switch to the RSG mass-loss rate as soon as a certain fraction of the envelope becomes convective and contains recombined hydrogen, it is applied for these massive models already during the YSG phase. The mass-loss rates from \citet{yan1:23} increase rapidly with luminosity, preventing these stars from expanding and reaching temperatures as low as those of observed RSGs.  The applied mass-loss rates during the cool supergiant phase effectively remove the H-rich layers from these massive stars, leaving them with surface H-abundances typically below $X_\mathrm{H}\lesssim0.3$ by the end of their lives. Additionally, since we use the locally working superadiabatic reduction method in MESA instead of MLT++, these stars can remain inflated even with such low surface H-abundances instead of contracting and becoming WR stars. The resulting overproduction of YSGs is evident in Fig.~\ref{fig:csg_comp}: the model predicts 23 YSGs in the LMC and 16 in the SMC beyond the HD limit ($\log(L/\lsun)\gtrsim5.5$ \citep{dav1:18}. However, no star is seen in this region of the HRD in any of these galaxies, nor in any other galaxies with a similar metallicity \citep{sch1:25}, underscoring this tension.

    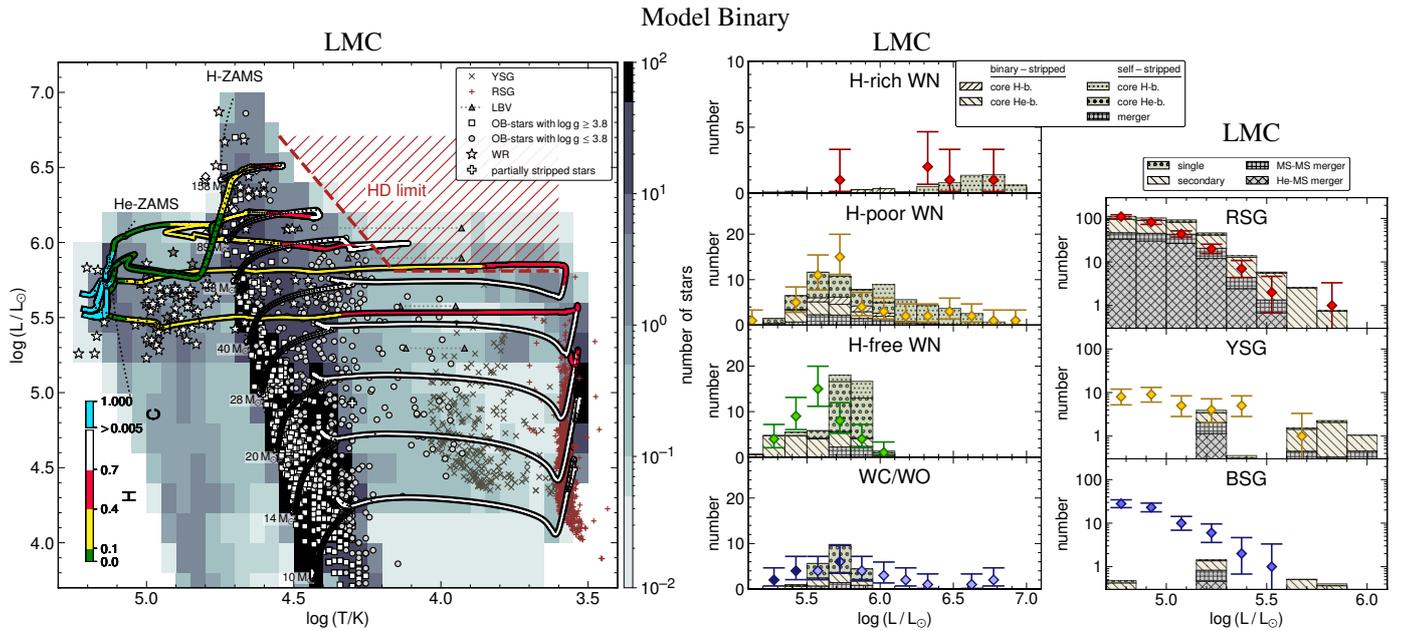
\begin{figure*}[tbhp]
        \centering
        \begin{tikzpicture}[scale=1]
            \node[anchor=south] at (-6.7, 0) {\includegraphics[trim={0.cm 0.75cm 0cm 1.2cm},clip,height=0.31\textheight]{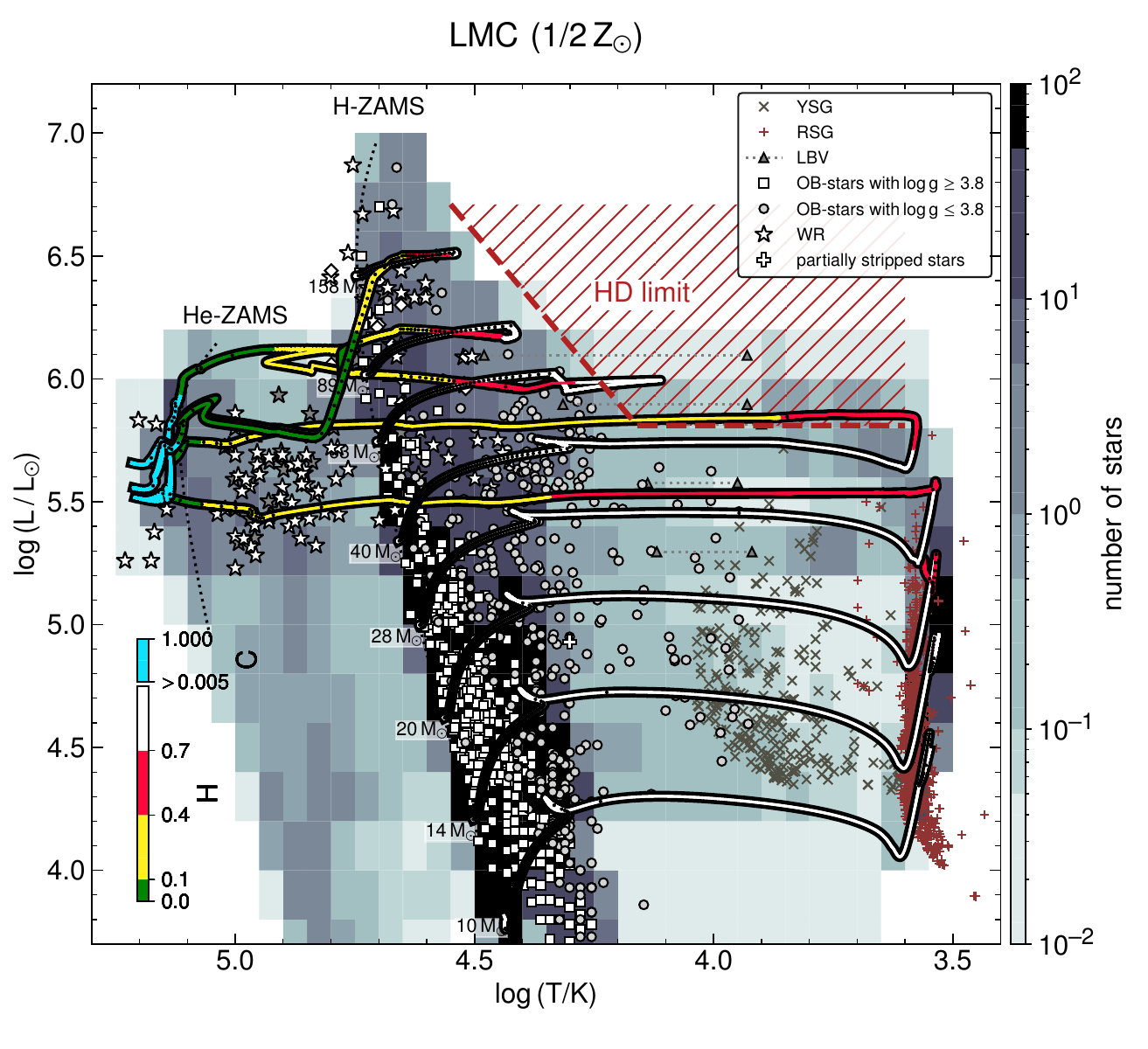}};
            \node at (4.48-11.2, 7.9) {LMC};
            \node[anchor=south] at (0.85+0.1+1.4, 0) {\includegraphics[trim={0.cm 0.75cm 0cm 0cm},clip,height=0.32\textheight]{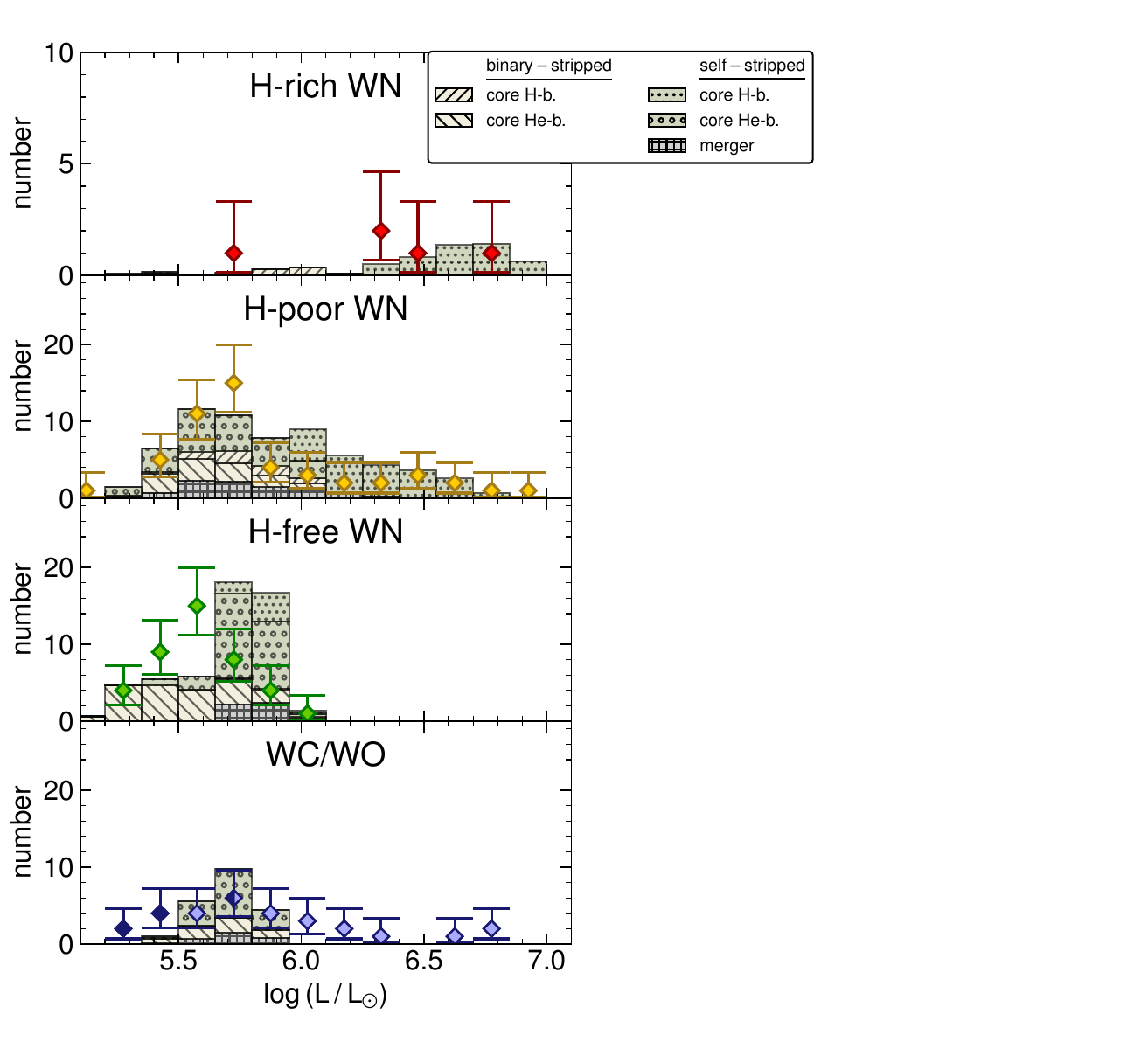}};
            \node at (-2+0.05, 8.2) {Model Binary}; 
            \node at (0.4+0.1, 7.9) {LMC};
            \node[anchor=south] at (4.6+0.1, 0) {\includegraphics[trim={0.cm 0.75cm 0cm 4.cm},clip,height=0.252\textheight]{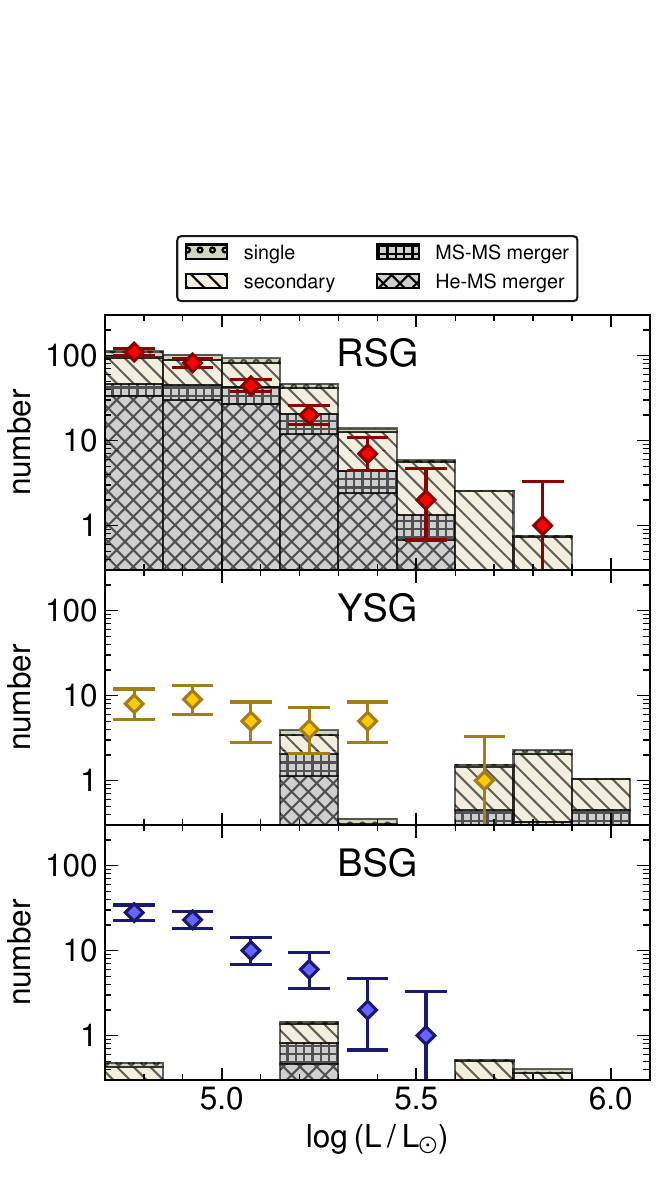}};
            \node at (5.+0.1, 6.7) {LMC};
        \end{tikzpicture}
        \caption{Hertzsprung-Russell diagram of the LMC population (left), luminosity distributions of the WR subtypes (center), and luminosity distribution of the cool supergiants (right) compared to the model predictions of our Model Binary.}
        \label{fig:binary}
    \end{figure*}
    
    \subsection{The role of binarity in stellar populations}
    \label{sec:binary}

    Large multi-epoch observing campaigns in the Galaxy, LMC, and SMC have established that most massive stars belong to a close binary system \citep[e.g.,][and references therein]{san1:25}. In binary systems, interactions such as mass transfer or common-envelope evolution can drastically alter the stars’ subsequent evolution and ultimate fate. \citet{pau1:22} computed an extensive grid of binary evolution models and concluded that the majority of WR stars in the LMC are likely formed via mass transfer. However, their models neglected Eddington-limit induced mass ejections and adopted the RSG mass-loss recipe from \citet{nie1:90} with a $Z$ dependence, which effectively prevented most single stars from evolving into WR stars (cf., Appendix~\ref{sec:rsg}). Conversely, \citet{she1:20} questioned whether binary interaction is the dominant WR formation channel at low $Z$, noting that the observed WR binary fraction appears independent of metallicity at roughly $30\%\text{\,--\,}40\%$ \citep[e.g.,][]{huc1:01,foe1:03,des1:24}.
    
    In this work, we assess the role of binary interactions across different stellar groups using our new Eddington-limit induced mass ejection prescription. Instead of calculating new binary model grids, which is computationally expensive, we combine our single-star models with the binary evolution models of \citet{pau1:22} and \citet{mar2:16} (for more details, see Sect.~\ref{sec:binary_pop}). The resulting predictions for the number of stars in each stellar group are listed in Table~\ref{tab:number_of_stars}.

    \subsubsection{The OB and VMS population}
    
    An HRD comparing the synthetic binary population at LMC metallicity with observations is shown in the left panel of Fig.~\ref{fig:binary}. Overall, it is quite similar to the single-star population (Fig.\ref{fig:hrd_lbv}), apart from the appearance of a He-star population located below the WR stars. The predicted number of O stars is about 3200, a factor of 1.2 times higher than in the single-star models. This number is sensitive to both the assumed post-interaction evolution of primaries, secondaries, and merger products, as well as the way one counts the stars that are seen by an observer. Given that the difference remains within a few tens of percent, we consider our models to still provide a good representation of the LMC population. Interestingly, because the model weights are assigned according to the IMF of the primary mass and the secondaries are assumed to follow a uniform mass-ratio distribution, the inclusion of binaries reduces the predicted number of VMSs to 19, bringing the models at the high mass end into closer agreement with observations.

    \subsubsection{The WR population}

    The LMC's synthetic binary population contains 120 WR stars, which is remarkably similar to the single-star case. In the central panel of Fig.~\ref{fig:binary}, we compare the luminosity distribution of the observed WR subtypes to our model predictions. Different hatch patterns indicate WR stars formed via binary interaction (stable mass-transfer) or self-stripping (including non-interacting binaries and mergers). Caution is needed when interpreting WR stars brighter than $\log(L/\lsun)\gtrsim6.1$, as in our models, these can only form through self-stripping. This limitation arises because the adopted binary grids cover initial primary masses only up to $M_\mathrm{1,\,ini}=89\,\msun$, meaning that interactions in more massive systems with orbits short enough to initiate mass transfer ($P_\mathrm{orb,\,ini}\lesssim\SI{50}{d}$) are not properly modeled.

    The luminosity distribution of H-poor WN stars is well reproduced. Interestingly, most of these stars ($\approx80\%$) originate from non-interacting binaries or mergers. In contrast, in the H-free WN population, the self-stripped fraction drops to $65\%$, particularly at low luminosities where binary interactions dominate. This is expected since single stars must first remove their extended H-rich envelopes before entering a WR phase, whereas binary mass transfer can efficiently strip most of the H-rich layers regardless of initial mass, allowing stars to more easily evolve into H-free WNs, especially at lower initial masses.

    In the H-free WN luminosity distribution, a distinct peak at $\log(L/\lsun)\approx5.8$ appears, which is dominated by self-stripped stars and not visible in the observed LMC's WR population. This feature is a by-product of our Eddington-limit induced mass ejection treatment. By limiting the maximum inflation, all stars with $M_\mathrm{ini}\gtrsim89\,\msun$ end their lives in the same HRD region (cf. the $89\,\msun$ and $150\,\msun$ tracks in Fig.~\ref{fig:binary}, left panel), leading to an overpopulation. Since binary evolution is not modeled for the most massive stars, this peak may be an artifact of our simplified approach to generating a binary population and might vanish in properly calculated binary evolution models using our setup.

    For WC/WO stars, our model predicts 16 objects compared to the 26 observed. Since we are only calculating our models until core He depletion, it might be that we accidentally cut off a part of this population. However, given that core C burning takes only about $10\%$ of the time a star spends core He burning, this would marginally increase the numbers and might only help explain the faint WO stars seen in the LMC.

    Above, we have only discussed the fraction of WR stars formed through binary interactions. This is not equivalent to the binary fraction of WR stars. \citet{sch1:08} observed that only $\approx30\%$ of WN stars in the LMC are in binaries with $P\lesssim\SI{200}{d}$. Since we do not know the current period for all systems, we try to estimate the binary fraction using the initial orbital period. Using this approach, we obtain that $43\%$ of the WN stars should be in a binary system with $P\lesssim\SI{200}{d}$. The discrepancy in the binary fractions, or at least a large part of it, can be explained by observational biases that were not taken into account when reporting the observed WR binary fraction. Overall, this agreement provides additional support for our model, which lifts parts of the long-standing question of how low-$Z$ WR stars form.
    
    \subsubsection{The cool supergiant population}

    Binary interaction and binary evolution are often associated with mass transfer, which is considered a shortcut to forming WR stars at the expense of skipping the RSG phase and thus causes there to be fewer RSGs in a binary population. In contrast to these expectations, our binary population contains 270 RSGs, which is slightly more stars than in the single-star population. The RSGs in our population originate both from mergers and ejected secondaries. This can be seen in the right panel of Fig.~\ref{fig:binary}, which illustrates the luminosity distribution of the cool supergiant population, separated into contributions from non-interacting binaries, secondaries, main-sequence mergers, and post-main-sequence mergers. The fraction of RSGs originating from mergers decreases with increasing luminosity: while $\approx25\%$ of the RSGs at $\log(L/\lsun)\lesssim5.4$ are merger products, this fraction drops sharply below $<5\%$ at higher luminosities. This behavior reflects the stability criteria for mass transfer adopted in our binary grids: lower-mass donors often undergo unstable mass transfer, whereas mass transfer from higher-mass donors is typically stable, thus leading to fewer mergers at higher initial masses.

    For the BSG and YSG populations, our models predict the same deficit as seen in the single-star population. This is expected, since we approximate mergers with our single-star models. However, the majority of mergers in our models ($\approx85\%$) occur after the primary has finished core-H burning. A merger of a post-main-sequence star and a main-sequence companion has a different core-to-envelope ratio, drastically impacting a star's further evolution. He-MS mergers are expected to evolve differently from ``normal'' single stars and spend more or even all time as BSGs or YSGs instead of as RSGs \citep[e.g.,][]{pod1:90,van1:13,jus1:14,sch1:25}. This effect could also help reduce the modest RSG excess we see in the synthetic population. This hypothesis is further supported by recent BLOeM campaign results in the SMC \citep{she1:24}, which indicate that nearly all BAF supergiants are apparently single \citep{pat1:25}, a fraction significantly lower than the ones measured for their main-sequence progenitors \citep{san1:25,vil1:25}. This suggests that the BAF population may contain a large fraction of merger products.

    We caution that these results and the discussion are based on several simplifying assumptions, including (i) all systems are disrupted after the primary’s death, (ii) our merger mass estimates, and (iii) the approximation of mergers by single-star tracks. Changing one of these could substantially affect the predicted cool supergiant population and its luminosity distribution. 
    
    \subsection{Comparison to other LBV models}

    The literature on LBV mass-loss prescriptions is sparse, limiting the scope for detailed comparisons with other implementations. \citet{gra1:21} performed time-dependent hydrodynamic stellar evolution calculations of a $73\,\msun$ model to demonstrate, on the example of AG\,Car, that LBV eruptions are linked to opacity and inflated envelopes near the Eddington limit. Their approach reproduces both the quiescent and outburst phases of LBVs, but is computationally expensive, taking over 100\,000 timesteps for the LBV phase alone. Nevertheless, we can roughly compare the physical properties of their Galactic model with our $71\,\msun$ LMC model. In the hydrodynamic models, the transition from quiescence to outburst occurs at $T=\SIrange{20}{25}{kK}$, with mass-loss rates of $\log(\dot{M}_\mathrm{LBV,\,Gras}/\msunpyr)=-2.9\text{\,\,to}-3.3$ during the outburst. Following our Eddington-limit induced mass ejection prescription, our $71\,\msun$ model enters the rapid mass ejection episode at $T=\SI{22}{kK}$ and reaches an average $\log(\dot{M}_\mathrm{eject}/\msunpyr)\approx-3.1$, which is in agreement with the model from \citet{gra1:21}.

    In the time-dependent hydrodynamic model, the LBV eruptions recur every $12$ years. Our prescription treats the Eddington-limit induced mass ejections, and thus a LBV phase, as a continuous outflow, which may underestimate the actual time spent in the LBV state. However, because the LBV phase in our models proceeds on a timescale comparable to the thermal timescale, even a tenfold increase in duration with intervening quiescent phases would have little effect on subsequent stellar evolution and would only impact the predicted number of LBVs.

    \citet{che1:24} used the MESA code to estimate the energy available in layers with super-Eddington luminosities that could be used to unbind the stellar envelope. They present stellar models at solar and SMC metallicity. In their prescription, the fraction of excess energy available for envelope ejection is treated as a free parameter between $0$ and $1$. Applying this formalism can yield mass-loss rates up to $\log(\dot{M}_\mathrm{LBV,\,Cheng}/\msunpyr)\approx-2$, comparable to our models. 
    
    Their Galactic-metallicity models using the maximum possible ejection efficiency exhibit a bimodal behavior: stars with $M_\mathrm{ini}=25\,\msun\text{\,--\,}60\,\msun$ undergo eruptive mass loss during the YSG phase, preventing them from evolving into RSGs (even at luminosities where RSGs are observed) while stars with $M_\mathrm{ini}\gtrsim70\,\msun$ lose mass too early, turning back before reaching the S\,Dor instability strip. In their Galactic models with lower ejection efficiencies as well as in their SMC-metallicity models, the eruptions are insufficient to remove most of the H-rich envelope, allowing the stars to remain beyond the HD limit for extended periods. This is in contrast to both our models and those of \citet{gra1:21}, which do not spend a significant fraction of their evolution beyond the HD limit except during Eddington-limit induced mass ejections.
        
    \section{Summary}
    \label{sec:conclusions}

    In this work, we have presented a new mass-loss prescription to account for mass ejections during the evolution of luminous massive stars approaching the Eddington limit, such as those expected during an LBV phase. Within our models, the mass ejections are triggered when stellar models inflate their envelope as a consequence of reaching the Eddington limit beyond a threshold value. The HD limit and the WR populations of the LMC and SMC were used as benchmarks to calibrate the maximum inflation a star can have before going through a mass ejection episode, thus effectively removing inflation.

    From our models, Eddington-limit induced mass ejections can be triggered during two main phases. Stars more massive than $M_\mathrm{ini}\gtrsim50\,\msun$ enter an LBV-like phase during the main sequence. Before the mass ejection episode, they have mass-loss rates comparable to those of the early O supergiants, while after the mass ejections, they have mass-loss rates comparable to the WNh stars; both  were observed in the same region of the HRD. Stars with $M_\mathrm{ini}\approx25\,\msun\text{\,--\,}40\,\msun$ evolve first into RSGs. As soon as they expel their H-rich envelope, radiation pressure dominates in the envelope, which is associated with envelope inflation. After the RSG phase, these stars experience short mass ejections in a region similar to those of the faintest observed LBVs.

    Our new Eddington-limit induced mass ejection model exhibits only a slight $Z$ dependence, as the opacity governs how much a star inflates. This $Z$ dependence is evident when comparing the LMC and SMC populations, although it remains consistent with observed data. Due to the weak $Z$ dependence, one would still expect that at very low metallicities, only a few stars would reside beyond the HD limit. These stars would be observable as cool supergiants rather than RSGs.

    Despite their simplicity, the models using our new physical setup successfully replicate many of the features observed in stellar populations, even at low $Z$, and they provide a definitive improvement compared to models without Eddington-limit induced mass ejections. Within the models, the Eddington-limit induced mass ejections occur in regions of the HRD where LBVs are observed. These mass ejections explain the lack of stars beyond the HD limit, including the upper limit of RSG luminosities. Furthermore, the number of observed O-type stars, WRs, and RSGs in the LMC and SMC can be matched. When using strong $Z$-independent RSG winds, such as those from \citet{yan1:23}, we can explain the faintest observed single WR stars; however, it is only when we use the Eddington-limit induced mass ejections that we can reproduce the observed low-$Z$ WO-type stars in the SMC and IC\,1613. 

    We combined our single-star models with large detailed grids of binary evolution models at LMC metallicity. According to our binary population, binary interactions still play a fundamental role in forming WR stars, especially for the H-free WN stars. However, a large fraction of the WR stars can be formed by stars in non-interacting binaries and mergers. One of the limitations of our Eddington-limit induced mass ejection model is that all stars $\gtrsim150\,\msun$ converge to the same track, leading to an excess of luminous H-free WR stars. Our binary population assumes that all stars are born in binary systems, with 70\% of the O-stars found in close binaries where they can interact. Interestingly, our new model predicts that about less than $40\%$ of the WR stars are in close binary systems, which is in agreement with observations. This further underlines the importance of Eddington-limit induced mass ejections in stellar populations.

    \begin{acknowledgements}
        DP acknowledges financial support from the FWO in the form of a junior postdoctoral fellowship No. 1256225N. AP acknowledges support from the FWO under grant agreement No. 11M8325N (PhD Fellowship), and K209924N, K223124N, K1A4925N (Travel Grants). PM acknowledges support from the European Research Council (ERC) under the European Union’s Horizon 2020 research and innovation programme (grant agreement No. 101165213/Star-Grasp), and from the Fonds Wetenschappelijk Onderzoek (FWO) senior postdoctoral fellowship number 12ZY523N.
    \end{acknowledgements}

 	\bibliographystyle{aa}                                                         
 	\bibliography{astro}

 \clearpage

   \begin{appendix}
        \section{Additional information on the assumed input physics}
        \label{app:MESA}
            Convection is modeled using the Ledoux criterion and standard mixing length theory, following \citet{cox1:68}, with a mixing length parameter $\alpha_\mathrm{mlt}=1.5$. Thermohaline mixing is included with an efficiency factor of $\alpha_\mathrm{th}=1$ \citep{kip1:80}. Rotational mixing in MESA is included as a diffusive process. Following \citet{heg1:00}, we include the effects of dynamical and secular shear instabilities, the Goldreich-Schubert-Ricke instability, and Eddington-Sweet circulations. The efficiency factors for rotational mixing are calibrated according to \citet{bro1:11}, with values set to $f_c=1/30$ and $f_\mu=0.1$. 
            
            The initial chemical compositions were as follows: for H and He, we used $X_\mathrm{H}=0.7343$ and $X_\mathrm{He}=0.2594$ in the LMC models, and $X_\mathrm{H}=0.7452$ and $X_\mathrm{He}=0.2522$ in the SMC models. For C, N, O, Ne, Na, Mg, Al, Si, S, Cl, Ar, Ca, Cr, Fe, and Ni, we adopted the recent baseline abundances for the LMC and SMC as reported by \citet{vin2:23}. For elements not specified in these abundances, we assumed they can be scaled down from the solar abundances provided by \citet{mag1:22}. The resulting metallicities for our LMC and SMC grids were $Z_\mathrm{LMC}=0.006177$ and $Z_\mathrm{SMC}=0.0024386$, respectively. For reference, the solar metallicity value of \citet{mag1:22} is $\zsun=0.012823$.
        
            To properly model the nuclear reactions of massive stars, we use a nuclear network inspired by the stellar evolution code STERN \citep{heg1:00,jin1:24}. We include 53 reaction rates for the p-p chains, CNO, Ne-Na, and Mg-Al burning, which are important already during core hydrogen and helium burning in VMSs. 
            
        \section{Calculating statistical weights for the population synthesis}
        \subsection{Single star population}
        \label{app:single_star}
            The total number of stars in a star population can be  calculated as
            \begin{equation}
                N_\mathrm{tot} = \int_{0.01}^{400}A\,\xi(m)dm\approx A \sum_{i=1}^n\xi(m_i)\Delta m_i\,,
            \end{equation}    
            where $m_i$ is the initial mass of a stellar model, $\Delta m_i$ is the difference in the initial mass between the successive stellar models (which is not constant, given the logarithmic spacing of our grid), $n$ is the maximum number of models in our grid, $A$ the normalization factor, and $\xi(m)$ the \citet{kro1:01} IMF given by
            \begin{equation}
                \xi(m)=\begin{cases}
                    \left(\dfrac{m}{0.08\,\msun}\right)^{-0.3} & 0.01<m<0.08\\
                    \left(\dfrac{m}{0.08\,\msun}\right)^{-1.3} & 0.08\leq m<0.5\\
                    \left(\dfrac{0.5\,\msun}{0.08\,\msun}\right)^{-1.3}\left(\dfrac{m}{0.5\,\msun}\right)^{-2.3} & 0.5\leq m\\
                \end{cases}.
                \label{eq:imf}
            \end{equation}
            Each stellar track is split into time steps of $\Delta t =\SI{1000}{yr}$ and it is assumed that both galaxies had constant star-formation rates (SFRs) with $\mathrm{SFR}_\mathrm{LMC}=0.1\,\msunpyr$ in the LMC \citep[corrected for a Kroupa IMF]{har1:09} and ${\mathrm{SFR}_\mathrm{SMC}=0.05\,\msunpyr}$ in the SMC \citep{rub1:18} within the last $\SI{27.5}{Myr}$ (i.e., the age of our oldest model). The predicted number of stars within an initial mass interval $[m,m+\Delta m]$ and age interval $[t,t+\Delta t]$ after a simulation time of $T_\mathrm{sim}$ is 
            \begin{equation}
                \Delta N = A\,\xi(m)\,\Delta m\,\Delta t/T_\mathrm{sim}.
            \end{equation}
            To calculate the normalization factor $A$, we use the condition that the total mass used to form stars within the simulation time is given by
            \begin{equation}
                M_\mathrm{tot}=\mathrm{SFR}\,\cdot T_\mathrm{sim} = \int_{0.01}^{400}m'A\,\xi(m')\mathrm{d}m'.
            \end{equation}
            Hence, the number of stars per initial mass bin $\Delta m$ per age step $\Delta t$ (as long as $t+\Delta t$ is less than the age of a star with the corresponding initial mass) in our simulation is
            \begin{equation}
                \Delta N = \dfrac{\mathrm{SFR}\cdot\xi(m)\,\Delta m\,\Delta t}{\int_{0.01}^{400}m'\,\xi(m')\mathrm{d}m'}.
            \end{equation}
            
        \subsection{Binary star population}
        \label{app:binary_star}

            The total number of stars in a population consisting only of binary stars is given by 
            \begin{equation}
                N_\mathrm{tot} = \int_{0.01}^{400}A\,\xi(m_1)dm_1\int_{0}^{1}f(q)\,dq\approx A\,\sum_{i=1}^n\xi(m_{1,i})\Delta m_{1,i}\Delta q\,,
            \end{equation}
            where $\xi(m)$ is the \citet{kro1:01} IMF (see Eq.~\ref{eq:imf}), $q=m_2/m_1$ is the initial mass ratio, and $f(q)$ the assumed mass-ratio distribution. Here we assume a uniform mass-ratio distribution, yielding $f(q)=q^0=1$.

            Similar to the calculation of the single-star population, we can use the total mass enclosed in the stars to calculate the normalization factor by
            \begin{align}
                M_\mathrm{tot}=B\cdot\mathrm{SFR}\,\cdot T_\mathrm{sim} &= A\,\int_{0.01}^{400}(m_1+q\,m_1)\,\xi(m_1)f(q)dq\,\mathrm{d}m_1. \nonumber \\
                 &= 1.5\,A\cdot\int_{0.01}^{400}m_1\,\xi(m_1)\,\mathrm{d}m_1.
            \end{align}
            The SFR used in this work is calculated using theoretical isochrones, photometry, and some assumptions to correct for the intrinsic distributions, such as the IMF. To correct the SFR of the LMC \citep{har1:09} for an intrinsic binary fraction of 100\%, we need to employ a correction factor of $B=2$.

            For the initial orbital periods, we assume all stars are born in binaries with $P_\mathrm{orb,\,ini}=\SI{1}{d}\text{\,--\,}\SI{100}{yr}$. Assuming that this range encompasses all possible binary configurations, the probability of finding a star in a binary whose initial period was in the interval  $[\log P, \log P+\Delta\log P]$ is
            \begin{equation}
                p(\log P,\log P+\Delta\log P)=\dfrac{\int_{\log P}^{\log P+\Delta\log P}\pi(\log P')d\log P'}{\int_0^{4.55}\,\pi(\log P'')\,d\log P''}.
            \end{equation}
            Following \citet{san2:12}, we adopt the intrinsic period distribution $\pi(\log P) = (\log P)^{-0.55}$.

            Analogous to the single-star case, the number of binaries formed per initial initial primary mass interval $[m_1,m_1+\Delta m_1]$, per time interval $[t,t+\Delta t]$, per initial mass ratio interval $[q+\Delta q]$, and per logarithmic initial orbital period interval $[\log P, \log P+\Delta\log P]$ is given by
            \begin{equation}
                \Delta N = \dfrac{2\cdot\,\mathrm{SFR}\,\xi(m_\mathrm{1})\Delta m_\mathrm{1}\,\Delta q\,\int_{\log P}^{\log P+\Delta\log P}\pi(\log P')d\log P' \,\Delta t}{1.5\int_{0.01}^{400}m_1'\xi(m_1')dm_1'\int_0^{4.55}\pi(\log P'')d\log P''}\,.
            \end{equation}

    \section{Detailed definitions of the distinct stellar groups}
    \label{app:groups}

            Wolf-Rayet stars exhibit optically thick winds, which lead to strong emission lines in their spectra. These emission features in the observations are used for classification. However, stellar evolution models do not output the spectrum of a star. To identify when a stellar model enters the WR phase, we follow the method described \citet{agu1:20} and calculate the optical depth of a stellar model using 
            \begin{equation}
                \tau=\dfrac{-\kappa\dot{M}}{4\pi R(\varv_\infty-\varv_0)}\,\ln\left(\dfrac{\varv_\infty}{\varv_0}\right)
            \end{equation}
            from \citet{lan1:89}, where $\kappa$ is approximated by the electron scattering opacity $\kappa_\mathrm{e}=0.2\,(1+X_\mathrm{H})\,\mathrm{cm^2\,g^{-1}}$, $\dot{M}$ is the mass-loss rate, $R$ is a stellar model's radius, $\varv_0$ is the expansion velocity at the photosphere, which is on the order of $\SI{20}{km\,s^{-1}}$, and $\varv_\infty$ is the terminal wind velocity, which is assumed to scale with the escape velocity $\varv_\mathrm{esc}$ as
            \begin{equation}
                \varv_\infty=n\cdot\varv_\mathrm{esc}=n\sqrt{\dfrac{2GM}{R}(1-\Gamma)}.
            \end{equation}
            Here, $\Gamma$ is the Eddington factor and $n$ is a scaling relation. While $n$ might vary during the evolution \citep[e.g., see][]{vin1:01}, we assume $n=1.3$, which provides a good approximation for hot O-type stars and WN-type stars \citep{vin1:01,gra1:11}. Following \citet{pau1:22}, who calibrated the $\tau$-criterion to the observed population of the WR stars in the LMC, we assume that a stellar model is in a WR phase as soon as $\tau\ge0.75$ and $\log(T/\mathrm{K})\ge4.6$ ($T\gtrsim40\,000\,\mathrm{K}$). 
                
           Wolf-Rayet stars have multiple subtypes, each characterized by emission lines from different elements, namely nitrogen, carbon, and oxygen. These subtypes are associated with distinct evolutionary phases and serve as excellent benchmarks for comparison of model predictions with observed populations. In this work, we define the following categories for the nitrogen-rich WN-type stars: H-rich WNs that have $X_\mathrm{H}>0.4$, H-poor WNs with $0.05< X_\mathrm{H}\leq0.4$, and H-free WNs are those WR stars with $X_\mathrm{H}\leq0.05$ and $X_\mathrm{C}<0.005$. While observational WC and WO-type stars are classified by their different strengths in carbon and oxygen lines, recent studies indicate that the different emission strengths are due to a temperature effect rather than an abundance effect \citep{aad1:22,san2:25}. Therefore, we do not differentiate between these two classes, and within our models, we define the WC/WO phase as soon as $X_\mathrm{C}\geq0.005$.
        
            The O-star phase in our models is assigned to all stars that are on the right side of the hydrogen zero age main sequence (H-ZAMS), have ${\log(T/\mathrm{K})\geq4.5}$ (${T\gtrsim31\,500\,\mathrm{K}}$), and an optically thin wind (i.e., $\tau<0.5$). The temperature cut corresponds to a spectral type of about O9.5. Note that we apply this temperature cut also to the observed dataset rather than using the spectral classification. 
        
            As VMSs, we define all stars with a luminosity above $\log(L/\lsun)\geq6.5$ (i.e., $M_\mathrm{ini}\gtrsim150\,\msun$). These stars, even though extremely rare in stellar populations, are dominating the integrated spectra of actively star-forming galaxies and are therefore of particular interest when trying to mimic real stellar populations. Note that the VMSs are also included in the WR and O-type stars sample. 
        
            To investigate how our synthetic populations perform, we additionally consider the evolutionary phases of BSG, YSG, and RSG. Comparing these stars provides valuable insights into the impact of mixing efficiencies as well as the performance of the assumed mass-loss rates. The RSG phase is assigned to every synthetic star that is cooler than ${\log(T/\mathrm{K})\leq3.68}$ (${T\lesssim4500\,\mathrm{K}}$), the YSG phase for all stars with ${3.68<\log(T/\mathrm{K}) \leq 3.87}$ (${4500\lesssim T\lesssim7500\,\mathrm{K}}$), and the BSG phase for stars with ${3.87<\log(T/\mathrm{K}) \leq 4.09}$ (${7500\lesssim T\lesssim12\,500\,\mathrm{K}}$). Since our grids only cover stars with initial masses above $10\,\msun$, they do not cover the complete cool supergiant population. As a consequence of varying physical assumptions, the maximum luminosity of the RSG of a star with an initial mass of $10\,\msun$ also changes. To ensure consistency, we apply a conservative luminosity cut at $\log(L/\lsun)\geq4.7$ for cool supergiants in both the synthetic and observed populations.
    
        \section{Additional variations of the input physics}
        \label{sec:add_changes}
        \subsection{Changing the RSG mass-loss rates}
        \label{sec:rsg}
        
        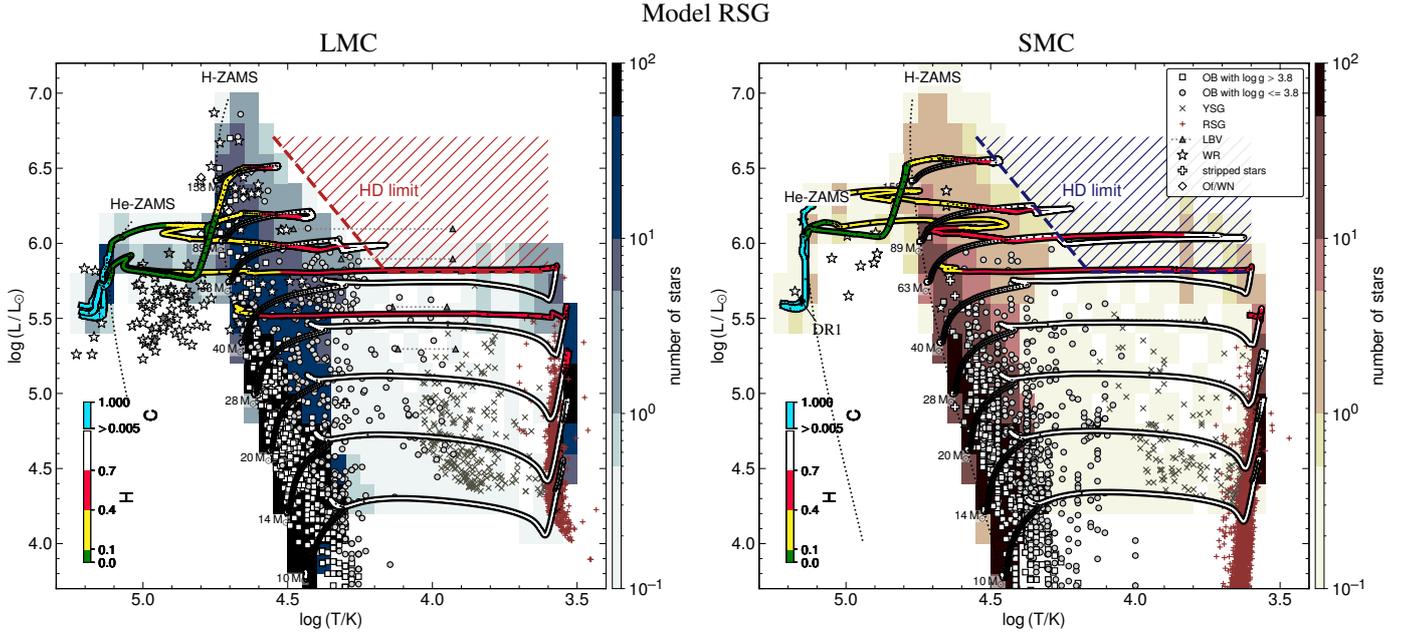
\begin{figure*}[tbhp]
            \centering
            \begin{tikzpicture}
                \node[anchor=south] at (0, 0) {\includegraphics[trim={0cm 0.75cm 0cm 1.2cm},clip,height=0.31\textheight]{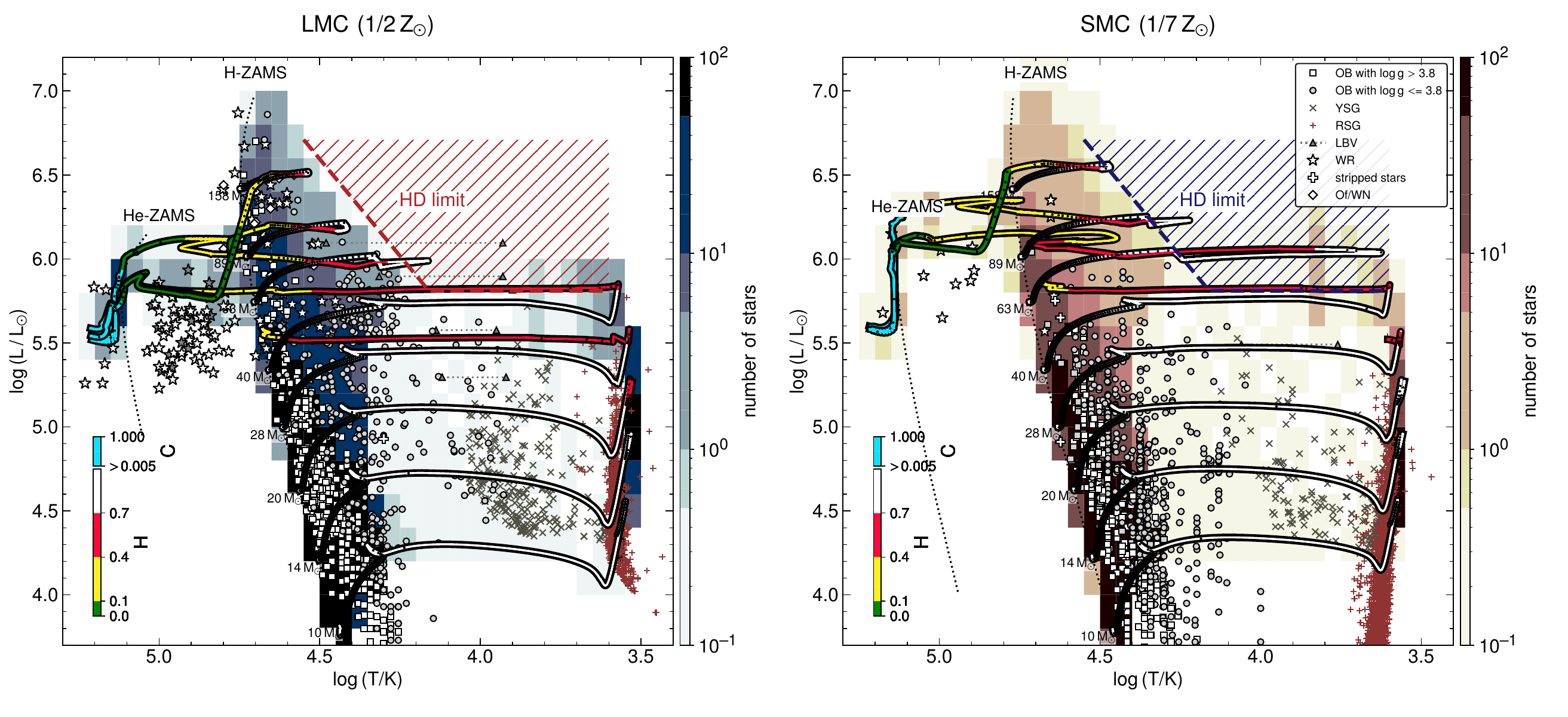}};
                \node at (0, 8.3) {Model RSG}; 
                \node at (-4.7, 7.9) {LMC};
                \node at (4.48, 7.9) {SMC};
                \node at (1.6, 4.1) {\fontsize{6}{14}\selectfont DR1};
                \draw[-latex] (1.45,4.2) -- (1.2,4.55);
            \end{tikzpicture}
            \caption{Hertzsprung-Russell diagrams of the massive star population (symbols) in the LMC (left), and SMC (right), compared to the synthetic population of the Model RSG (contours). The symbols and colors have the same meaning as in Fig.~\ref{fig:hrd_inflation}.}
            \label{fig:hrd_rsg}
        \end{figure*}

        One of the biggest uncertainties in modeling massive star evolution, besides the Eddington-limit induced mass ejections, is the mass lost via the winds of RSG. Comparing RSG mass-loss rates derived using different model assumptions (such as the launching mechanism of the wind) shows that the measured RSG mass-loss rate can differ by up to two orders of magnitude \citep[e.g.,][]{dec1:21,yan1:23,bea1:23, ant1:24, dec1:24}.
        These differences arise partly from the method of modeling itself \citep[e.g., see figure 13 in][]{ant1:24}. When comparing different recent mass-loss prescriptions \citep[e.g.,][]{bea1:23,yan1:23,dec1:24}, it becomes evident that most of them align within a factor of a few for luminosities above $\log(L/\lsun)\gtrsim5$ (i.e., the potential progenitors of WR stars) and that stronger mass-loss rates compared to the ``classic'' \citet{nie1:90} are observed. Still 
        in many publicly available stellar evolution grids, the mass-loss recipe of \citet{nie1:90} is commonly employed, sometimes even with a $Z$ dependence. Here, we present our Model RSG set, where we have switched from the \citet{yan1:23} mass-loss recipe to that of \citet{nie1:90}, but without any $Z$ dependence.
    
        The HRDs of the LMC and SMC populations compared to the synthetic Model RSG population are shown in Fig.~\ref{fig:hrd_rsg}. Since Eddington-limit induced mass ejections are still included, stellar models with $M_\mathrm{ini}\gtrsim50\,\msun$ continue to avoid the HD limit and evolve into WR stars. One can see that these stars only produce the most luminous WR stars in the population, as well as the majority of WC/WO-type stars. For stars with $M_\mathrm{ini}\approx25\,\msun\,\text{--}\,40\,\msun$, which lost their H-rich envelopes during the RSG phase in the Model Eject, most of them can no longer strip off their envelopes when employing the RSG mass-loss rates from \citet{nie1:90} and thus do not evolve into WR stars. This challenges stellar evolution in explaining the faintest observed single WR stars in both the LMC and SMC.
            
        In Fig.~\ref{fig:csg_comp}, we compare the luminosity distributions of BSG, YSG, and RSG in the LMC and SMC to those of the Model RSG population. It is evident that the high-luminosity bins ($\log(L/\lsun)\gtrsim5.5$) of the RSG contain too many stars. This is because these stars never lose their H-rich envelopes, and they spend all of their core He-burning lives as RSGs, contrary to the stars in Model Eject, which spend a significant fraction of core He burning as WR stars. The number of BSG and YSG is underpredicted at low luminosities ($\log(L/\lsun)\lesssim5.5$) and is overpredicted in the model at high luminosities ($\log(L/\lsun)\gtrsim5.5$). We conclude that in order to achieve a better agreement with the observed stellar populations, stronger RSG mass-loss rates are favored, as this seems to be the only way to explain the observed faintest single low-$Z$ WR stars.
        
        \subsection{More efficient semiconvective mixing}        
    
\begin{figure}[tbhp]
            \centering
            \begin{tikzpicture}
                \node[anchor=south] at (0, 0) {\includegraphics[trim={0cm 0.75cm 0cm 1.2cm},clip,height=0.31\textheight]{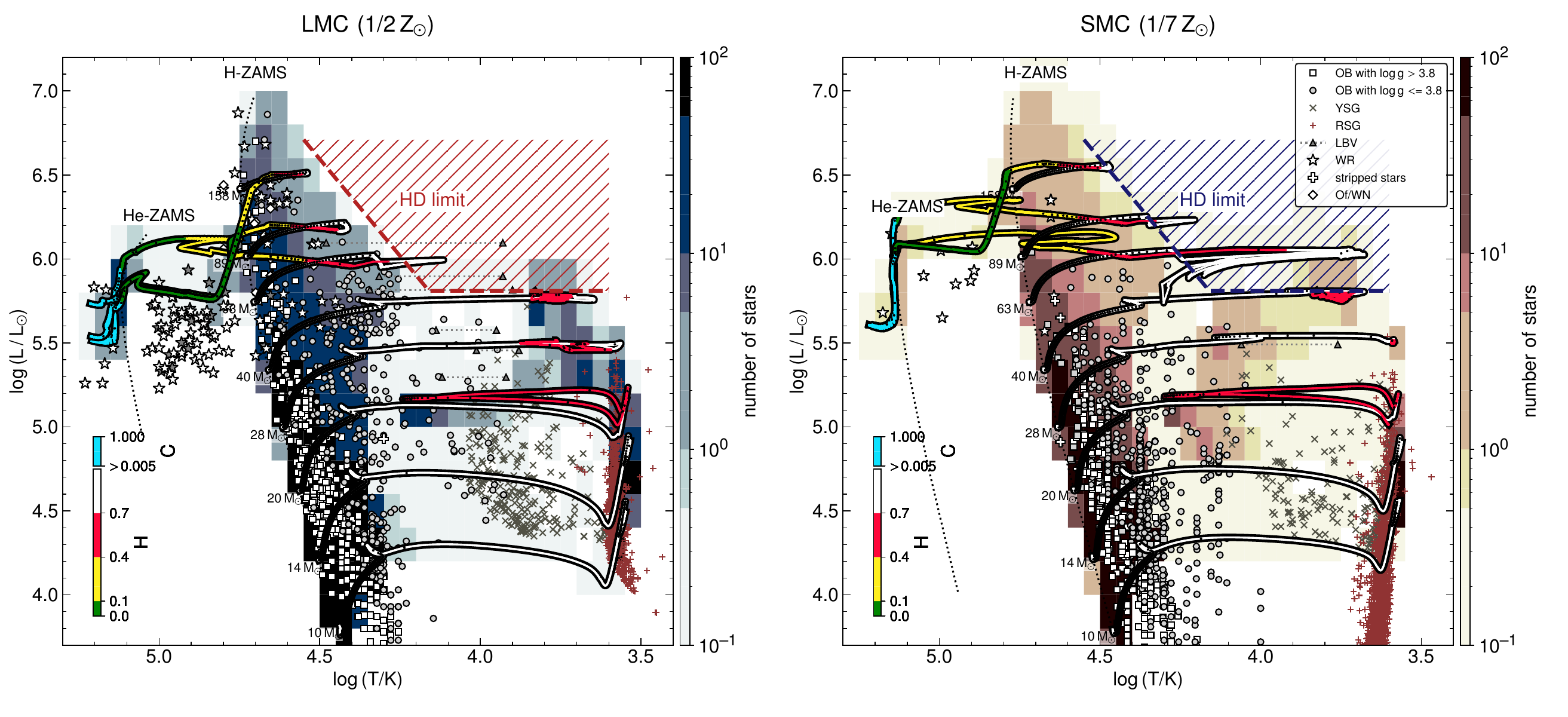}};
                \node at (0, 8.3) {Model sc10}; 
                \node at (-4.7, 7.9) {LMC};
                \node at (4.48, 7.9) {SMC};
                \node at (1.6, 4.1) {\fontsize{6}{14}\selectfont DR1};
                \draw[-latex] (1.45,4.2) -- (1.2,4.55);
            \end{tikzpicture}
            \caption{Hertzsprung-Russell diagrams of the massive star population (symbols) in the LMC (left) and SMC (right) compared to the synthetic population of the Model sc10 (contours). The symbols and colors have the same meaning as in Fig.~\ref{fig:hrd_inflation}.}
            \label{fig:hrd_sc10}
        \end{figure}
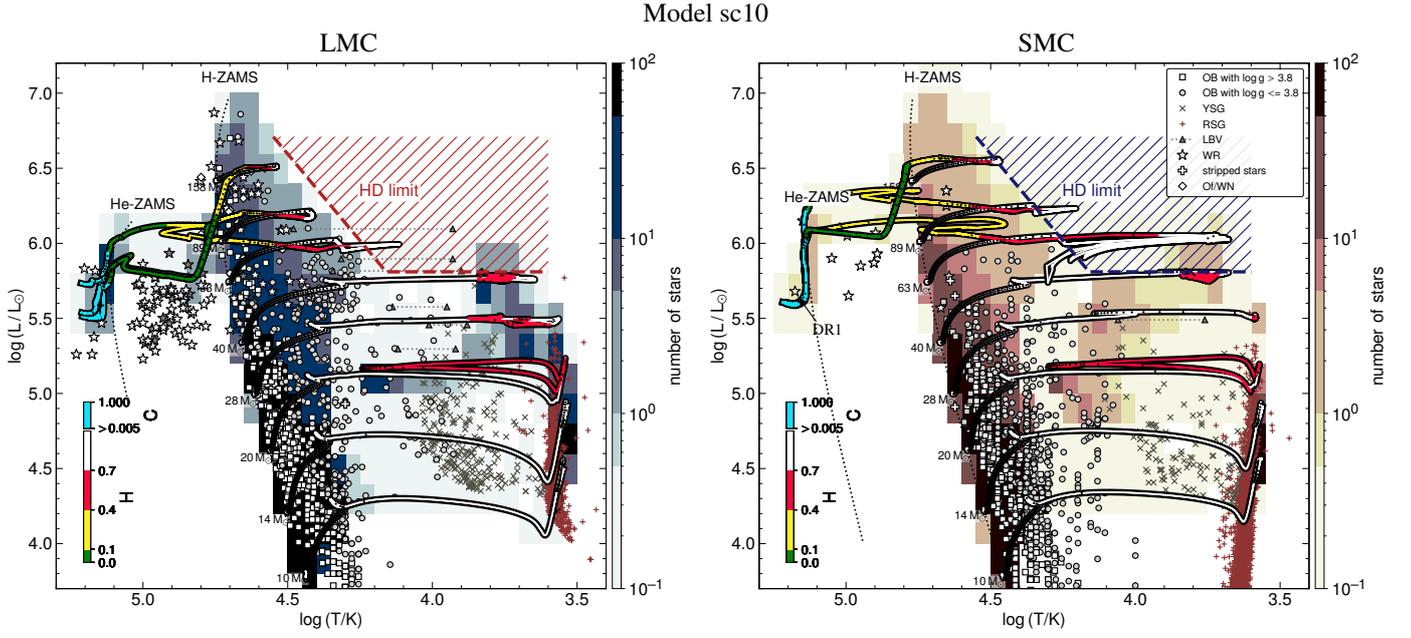

        All the aforementioned single-star models fail to reproduce the observed BSG and YSG populations in the LMC and SMC. Previous studies have shown that semiconvective mixing has a strong influence on the late evolutionary stages of massive stars, as it shortens the Hertzsprung gap, affects the He-core mass, and can lead to blue loops during which the star appears as a BSG or YSG \citep[e.g.,][]{sch1:19,gil1:19,kle1:20}. In Fig.~\ref{fig:hrd_sc10}, the HRD for Model sc10, which adopts the same physical setup as Model Eject but with a tenfold increase in the efficiency of semiconvective mixing, is shown. The impact on stars with initial masses $M_\mathrm{ini}\approx14\,\msun\text{\,--\,}40\,\msun$ is striking: instead of remaining in the RSG phase, these stars evolve through the expected blue loops.
    
        The luminosity distribution of cool supergiants for this model is shown in Fig.~\ref{fig:csg_comp}. While the blue-loop evolution successfully populates both the BSG and YSG regions, most of the predicted stars are found at $\log(L/\lsun)\gtrsim5.5$, a regime where observations show few to no objects. Moreover, the Model sc10 predicts more YSGs than BSGs, opposite to what is observed. Both discrepancies may stem from our adopted semiconvection and overshooting parameters. As shown by \citet{sch1:19}, stars at SMC metallicity only undergo blue loops for specific combinations of overshooting and semiconvection. It is still unclear whether these mixing parameters, such as overshooting, may vary with mass and what this implies for the cool supergiant population \citep[e.g.,][]{cla1:16,jer1:18,gil1:19}.
    
        A further issue of the Model sc10 is that the stars spend substantial time as BSGs or YSGs and thus do not experience strong RSG winds. This prevents them from shedding their H-rich envelopes and evolving into WR stars, which is again conflicting with observations. Interestingly, recent multi-epoch observations of BAF supergiants in the SMC suggest that most, maybe all, of them are apparently single \citep{pat1:25}. This raises the possibility that many observed BSGs and YSGs are in fact merger products or ejected accretors \citep{men1:24}. It has been shown that accretors that gain significant mass are not fully rejuvenated and exhibit different core-to-envelope ratios and internal structures \citep[e.g.,][]{kle1:22,ren1:23,pau1:23}, which can naturally lead to blue-loop evolution without invoking highly efficient semiconvective mixing. A similar evolutionary pathway is expected for post-main-sequence merger products \citep[e.g.,][]{van1:13,jus1:14}. Consequently, the inability of single-star models to match the observed BSG and YSG populations may not be a major concern.
        \FloatBarrier
        \onecolumn
        \section{Additional figures}
        \begin{figure}[h!]
            \centering
            \begin{tikzpicture}
                \node[anchor=south] at (0, 0) {\includegraphics[trim={0cm 0.75cm 0cm 1.2cm},clip,height=0.35\textheight]{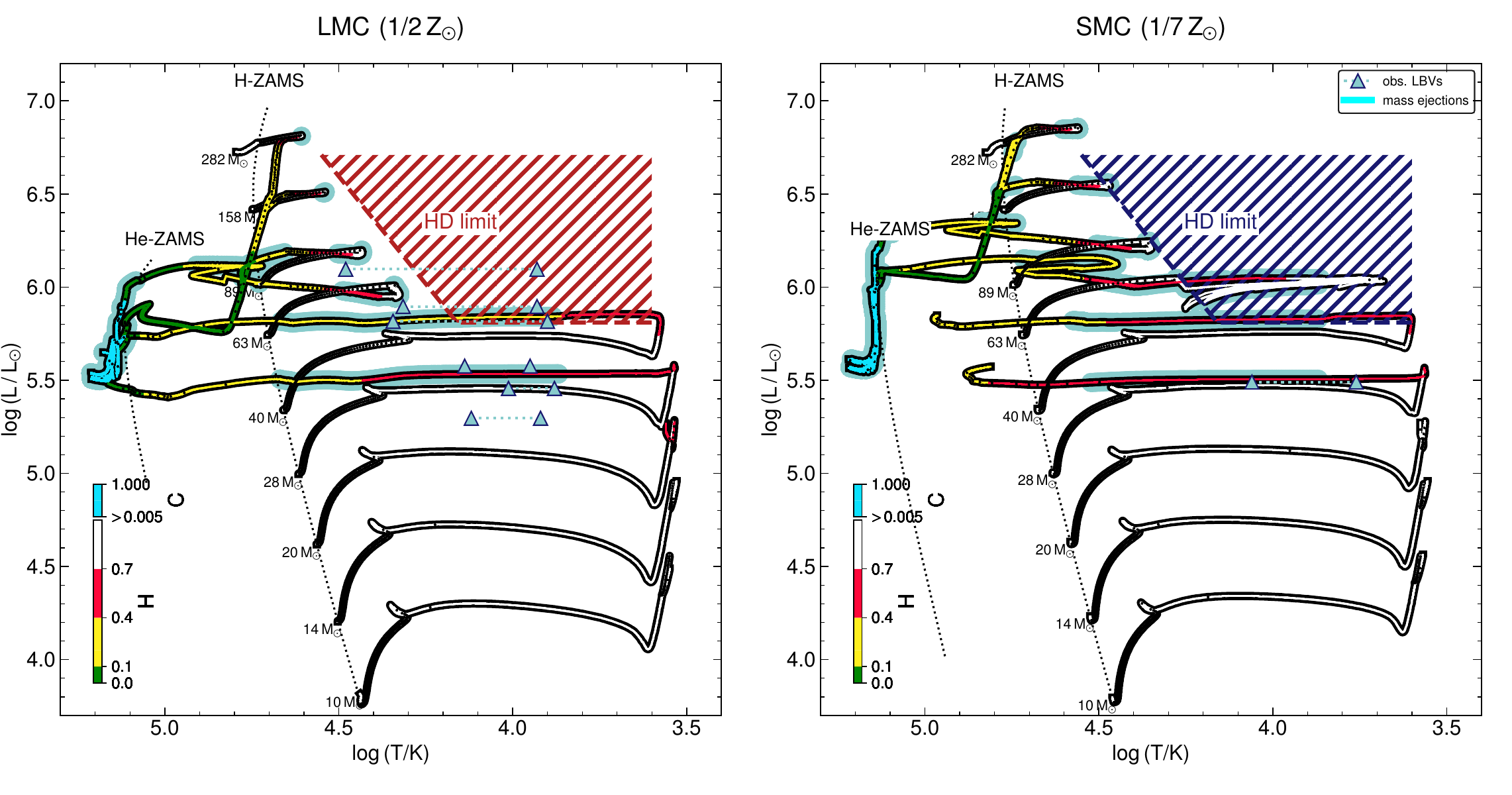}};
                \node at (0, 9.4) {Model Eject}; 
                \node at (-4.5, 8.9) {LMC};
                \node at (4.85, 8.9) {SMC};
            \end{tikzpicture}
            \caption{Hertzsprung-Russell diagrams showing the tracks of our Model Eject grid for LMC (left) and SMC metallicity (right). The Eddington-limit induced mass ejection episodes of our models are indicated by bold gray outlines around the tracks. Black dots on the tracks indicate equidistant timesteps of $\Delta t=\SI{30000}{yr}$. For comparison, the known LBVs of the LMC and SMC are shown as triangles during quiescence (left triangle) and outburst (right triangle) \citep{hum1:16,kal1:18}.}
            \label{fig:hrd_phases}
        \end{figure}
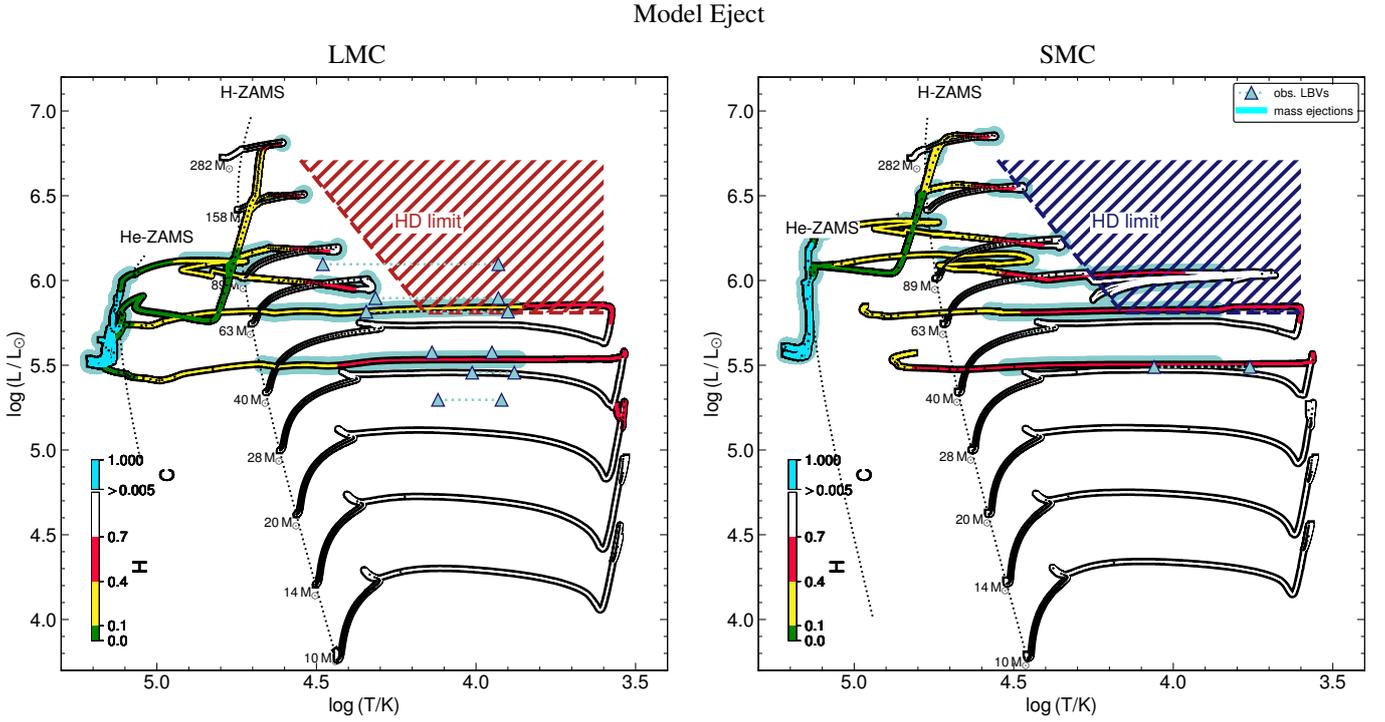
        
    \end{appendix}

\end{document}